\documentclass[twocolumn,showpacs,preprintnumbers,amsmath,amssymb,superscriptaddress,nofootinbib,english]{revtex4-1}
\usepackage{times,amsmath,amsfonts,amssymb,epstopdf}
\usepackage{graphicx}
\usepackage{dcolumn}
\usepackage{bm}
\usepackage{epsfig}
\usepackage{graphicx}
\usepackage{hyperref}
\usepackage[usenames]{color}
\usepackage{url}
\usepackage[normalem]{ulem}
\usepackage[T1]{fontenc}

\def\be{\begin{equation}}
\def\ee{\end{equation}}
\def\ben{\begin{eqnarray}}
\def\een{\end{eqnarray}}
\def\ba{\begin{array}}
\def\ea{\end{array}}

\newcommand{\bq}{\begin{eqnarray}}
\newcommand{\eq}{\end{eqnarray}}
\newcommand{\bes}{\begin{subequations}}
\newcommand{\ees}{\end{subequations}}

\begin{document}
\newcommand{\half}{{\textstyle\frac{1}{2}}}
\allowdisplaybreaks[3]
\def\triangledown{\nabla}
\def\grad3{\hat{\nabla}}
\def\a{\alpha}
\def\b{\beta}
\def\g{\gamma}\def\G{\Gamma}
\def\d{\delta}\def\D{\Delta}
\def\ep{\epsilon}
\def\et{\eta}
\def\z{\zeta}
\def\t{\theta}\def\T{\Theta}
\def\l{\lambda}\def\L{\Lambda}
\def\m{\mu}
\def\f{\phi}\def\F{\Phi}
\def\n{\nu}
\def\p{\psi}\def\P{\Psi}
\def\r{\rho}
\def\s{\sigma}\def\S{\Sigma}
\def\ta{\tau}
\def\x{\chi}
\def\o{\omega}\def\O{\Omega}
\def\k{\kappa}
\def\pa {\partial}
\def\ov{\over}
\def\br{\\}
\def\ud{\underline}

\newcommand\lsim{\mathrel{\rlap{\lower4pt\hbox{\hskip1pt$\sim$}}
    \raise1pt\hbox{$<$}}}
\newcommand\gsim{\mathrel{\rlap{\lower4pt\hbox{\hskip1pt$\sim$}}
    \raise1pt\hbox{$>$}}}
\newcommand\esim{\mathrel{\rlap{\raise2pt\hbox{\hskip0pt$\sim$}}
    \lower1pt\hbox{$-$}}}
\newcommand{\dpar}[2]{\frac{\partial #1}{\partial #2}}
\newcommand{\sdp}[2]{\frac{\partial ^2 #1}{\partial #2 ^2}}
\newcommand{\dtot}[2]{\frac{d #1}{d #2}}
\newcommand{\sdt}[2]{\frac{d ^2 #1}{d #2 ^2}}    

\title{Linear perturbations in Galileon gravity models}

\author{Alexandre Barreira}
\email[Electronic address: ]{a.m.r.barreira@durham.ac.uk}
\affiliation{Institute for Computational Cosmology, Department of Physics, Durham University, Durham DH1 3LE, U.K.}
\affiliation{Institute for Particle Physics Phenomenology, Department of Physics, Durham University, Durham DH1 3LE, U.K.}

\author{Baojiu Li}
\affiliation{Institute for Computational Cosmology, Department of Physics, Durham University, Durham DH1 3LE, U.K.}

\author{Carlton M. Baugh}
\affiliation{Institute for Computational Cosmology, Department of Physics, Durham University, Durham DH1 3LE, U.K.}

\author{Silvia Pascoli}
\affiliation{Institute for Particle Physics Phenomenology, Department of Physics, Durham University, Durham DH1 3LE, U.K.}

\begin{abstract}
We study the cosmology of Galileon modified gravity models in the linear perturbation regime. We derive the fully covariant and gauge invariant perturbed field equations using two different methods, which give consistent results, and solve them using a modified version of the {\tt CAMB} code. We find that, in addition to modifying the background expansion history and therefore shifting the positions of the acoustic peaks in the cosmic microwave background (CMB) power spectrum, the Galileon field can cluster strongly from early times, and causes the Weyl gravitational potential to grow, rather than decay, at late times. This leaves clear signatures in the low-$l$ CMB power spectrum through the modified integrated Sachs-Wolfe effect, strongly enhances the linear growth of matter density perturbations and makes distinctive predictions for other cosmological signals such as weak lensing and the power spectrum of density fluctuations. The quasi-static approximation is shown to work quite well from small to the near-horizon scales. We demonstrate that Galileon models display a rich phenomenology due to the large parameter space and the sensitive dependence of the model predictions on the Galileon parameters. Our results show that some Galileon models are already ruled out by present data and that future higher significance galaxy clustering, ISW and lensing measurements will place strong constraints on Galileon gravity.
\end{abstract} 
\maketitle

\section{Introduction}

The accumulated evidence for the present-day accelerated expansion of the Universe, driven by what is generically referred to as `dark energy', is now overwhelming \cite{Reid:2009xm, Komatsu:2010fb, Amanullah:2010vv}. The simplest explanation for the nature of dark energy is a simple cosmological constant but, despite the good agreement with the observational data so far, such an explanation is plagued with serious fine tuning and coincidence problems. This has motivated the proposal of alternative models to explain the observations, the majority of which fall into two classes. The first one assumes the existence of a dynamical dark energy field (often of scalar type) which dominates the energy density today and has a negative pressure to accelerate the Universe \cite{2006IJMPD..15.1753C, Li:2011sd}. The other considers that the standard law of gravity, general relativity, fails on cosmological scales and must be completed by modifications capable of accelerating the Universe \cite{Clifton:2011jh}. Models in the second class have attracted a lot of research interest recently, and significant progress has been made in both the theoretical modelling \cite{Brax:2011aw, Brax:2012gr} and numerical simulations \cite{Li:2011vk,Jennings:2012pt,Li:2012by}.

One notable example of a modified gravity model which has been the subject of many recent papers is the Galileon model \cite{PhysRevD.79.064036,PhysRevD.79.084003}. Here, the deviation from general relativity is mediated by a scalar field $\varphi$, dubbed the Galileon, whose Lagrangian is invariant under the Galilean shift symmetry $\partial_\mu\varphi \rightarrow \partial_\mu\varphi + b_\mu$ (hence the name), where $b_\mu$ is a constant vector. Such a field appears, for instance, as a brane-bending mode in the decoupling limit of the four-dimensional boundary effective action of the DGP braneworld model \cite{2000PhLB..485..208D, 2003JHEP...09..029L, 2004JHEP...06..059N} which was proposed well before the Galileon model. However, in spite of being theoretically appealing, the self-accelerating branch of the DGP model, which is of interest to the cosmological community, is plagued by the ghost problem \cite{1126-6708-2003-09-029, 1126-6708-2004-06-059, PhysRevD.73.044016} (i.e. there is not a well defined minimum energy). Taking the DGP model as inspiration, it was shown in \cite{PhysRevD.79.064036} that in four-dimensional Minkowski space there are only five Galilean invariant Lagrangians that lead to second-order field equations, despite containing highly nonlinear derivative self-couplings of the scalar field. The second-order nature of the equations of motion is crucial to avoid the presence of Ostrogradski ghosts \cite{Woodard:2006nt}. In \cite{PhysRevD.79.084003, Deffayet:2009mn}, it was shown how these Lagrangians could be generalised to curved spacetimes. These authors concluded that explicit couplings between the Galileon field derivatives and curvature tensors are needed to keep the equations of motion up to second-order. Such couplings however break the Galileon symmetry which is only a symmetry of the model in the limit of flat spacetime \cite{Hui:2012qt}. The couplings of the Galileon field and the curvature tensors in the equations of motion change the way in which spacetime responds to matter distributions, which is why the Galileon model is a subclass of modified gravity theories.

Since the equations of motion are kept up to second order, it means that the Galileon model is a subclass of the more general Horndeski theory \cite{Horndeski:1974wa, Deffayet:2011gz, Kobayashi:2011nu}. The Horndeski action is the most general single scalar field action one can write that yields only second order field equations of motion of the metric and scalar fields. Besides the Galileon model, it therefore encompasses simpler cases such as Quintessence, k-essence \cite{2006IJMPD..15.1753C} and $f(R)$ \cite{Sotiriou:2008rp, DeFelice:2010aj} models as well as other models which also involve derivative couplings of the scalar field that have recently generated some interest such as Kinetic Gravity Braiding \cite{Deffayet:2010qz, Pujolas:2011he, Kimura:2011td}, Fab-Four \cite{Charmousis:2011bf, Charmousis:2011ea, Bruneton:2012zk, Copeland:2012qf, Appleby:2012rx}, k-mouflage \cite{Babichev:2009ee} and others \cite{Kobayashi:2009wr, Kobayashi:2010wa, Leon:2012mt}. An important difference between the Galileon model and some other corners of Horndeski's general theory is that in the Galileon model there are no free functions since the fuctional form of the Lagrangian is fixed by the shift symmetry (see however \cite{DeFelice:2010nf}).

In any viable modified gravity theory, it is crucial that deviations from standard gravity get suppressed (or screened) in high matter-density regions where general relativity has been tested to high accuracy \cite{1990PhRvL..64..123D, 1999PhRvL..83.3585B}. In the case of Galileon gravity, such a screening is realised via the Vainshtein mechanism \cite{Vainshtein1972393}, which relies on the presence of the nonlinear derivative self-couplings of the Galileon field. Here, far away from gravitational sources, the nonlinear terms are subdominant and the Galileon field satisfies a linear Poisson equation (as the Newtonian potential), so that the extra (fifth) force mediated by it can be sizeable and proportional to standard gravity, effectively renormalising Newton's constant. Near the sources, on the other hand, the nonlinear terms become important, which strongly suppress the spatial variations of the Galileon field compared to that of the Newtonian potential and ensure that the extra force, which is the gradient of the Galileon field, is not felt on scales smaller than a given `Vainshtein radius'.  In certain respects, this is very similar to the chameleon screening \cite{PhysRevD.69.044026, PhysRevD.70.123518}, which operates for instance in $f(R)$ gravity models \cite{Sotiriou:2008rp, DeFelice:2010aj, Carroll:2004de, Brax:2008hh}. However, in the chameleon case the self-interaction of the scalar field depends on the field value (through a nonlinear interaction potential) rather than its derivatives, and the non-derivative coupling of the scalar field to matter makes its behaviour highly sensitive to the environmental matter density -- in high density regions the field value, rather than merely its gradient, becomes extremely small so that the extra force is suppressed.

It is therefore evident that one has to go beyond the local environment to look for possible deviations from general relativity and distinct signatures of the different modified gravity models. In particular, a promising way is to look at the cosmic expansion and the formation of structure in the Universe: different screening mechanisms in different modified gravity models can lead to very different predictions as to when, where and how the various cosmological observables are affected.

The effects of Galileon gravity models on the background cosmological expansion have already been studied in the literature in great detail \cite{Gannouji:2010au, PhysRevD.80.024037, DeFelice:2010pv, Nesseris:2010pc, Appleby:2011aa, PhysRevD.82.103015}. It has been shown that in these models there is  a stable de Sitter point that can be reached after the radiation and matter dominated eras, thus yielding a viable cosmological expansion history. Conditions to avoid the ghosts and other theoretical instabilities have also been derived by considering the linear perturbations \cite{DeFelice:2010pv, Appleby:2011aa}. 

To improve our understanding of the cosmological effects of Galileon gravity models and make direct comparisons with observational data, a proper investigation of the evolution of density fluctuations and formation of large-scale structure is necessary. Here, as an initial step, we consider the regime in which the density fluctuations are small such that their evolution is well described by linear perturbation theory. This regime is relevant for several important cosmological observables, such as the power spectrum of the cosmic microwave background (CMB) temperature fluctuations and its polarisations, the growth of matter density perturbations, the weak gravitational lensing of distant galaxies and the CMB map, and the integrated Sachs-Wolfe (ISW) effect and its cross correlation with the galaxy distribution. The rich information contained in this regime therefore warrants a detailed study of the Galileon effects, which is precisely the topic of this paper. The nonlinear regime of structure formation can in principle contain further interesting information, but its study is beyond the scope of the current paper.

The layout of this paper is as follows. We start by briefly presenting the Galileon model and the Galileon and metric field equations of motion in Section~\ref{The model}. The perturbation equations are derived and presented in a covariant and gauge invariant (CGI) way in Section \ref{The perturbation equations} using the method of $3+1$ decomposition. In Appendix~\ref{alternative_derivation} we present an alternative and considerably simpler derivation of the perturbation equations which is particularly suitable for the Galileon model as it takes advantage of the fact that the Lagrangian density is fixed by the Galilean shift invariance and that there are no derivatives higher than second order. We present and discuss the results for the CMB, lensing and linear matter power spectra in Section \ref{Results} which we obtain using a version of the {\tt CAMB} code \cite{camb_notes} which we have modified. In Section \ref{Results} we also discuss the time evolution of the gravitational potential, Galileon field perturbation and Galileon density contrast and the validity of the quasi-static limit. We conclude in Section \ref{Conclusion}.

Throughout this paper we will use the unit $c=1$ and metric convention $(+,-,-,-)$. Greek indices run over $0,1,2,3$ and we will use $8\pi G=\kappa=M^{-2}_{\rm Pl}$ interchangeably, where $G$ is Newton's constant and $M_{\rm Pl}$ is the reduced Planck mass.

\section{The model}

\label{The model}

The covariant uncoupled Galileon action can be written as \cite{PhysRevD.79.084003}

\bq\label{Galileon action}
&& S = \int d^4x\sqrt{-g} \left[ \frac{R}{16\pi G} - \frac{1}{2}\sum_{i=1}^5c_i\mathcal{L}_i - \mathcal{L}_m\right],
\eq
where $g$ is the determinant of the metric, $R$ is the Ricci scalar and $c_{1-5}$ are dimensionless constants.
The five covariant terms in the Lagrangian densities, which are fixed by the Galilean invariance in flat spacetime, $\partial_\mu\varphi \rightarrow \partial_\mu\varphi + b_\mu$, are given by
\bq\label{L's}
\mathcal{L}_1 &=& M^3\varphi, \nonumber \\
\mathcal{L}_2 &=& \nabla_\mu\varphi\nabla^\mu\varphi,  \nonumber \\
\mathcal{L}_3 &=& \frac{2}{M^3}\Box\varphi\nabla_\mu\varphi\nabla^\mu\varphi, \nonumber \\
\mathcal{L}_4 &=& \frac{1}{M^6}\nabla_\mu\varphi\nabla^\mu\varphi\left[ 2(\Box\varphi)^2 - 2(\nabla_\mu\nabla_\nu\varphi)(\nabla^\mu\nabla^\nu\varphi) \right. \nonumber \\
&&\left. -R\nabla_\mu\varphi\nabla^\mu\varphi/2\right], \nonumber \\
\mathcal{L}_5 &=&  \frac{1}{M^9}\nabla_\mu\varphi\nabla^\mu\varphi\left[ (\Box\varphi)^3 - 3(\Box\varphi)(\nabla_\mu\nabla_\nu\varphi)(\nabla^\mu\nabla^\nu\varphi) \right. \nonumber \\
&&\left. + 2(\nabla_\mu\nabla^\nu\varphi)(\nabla_\nu\nabla^\rho\varphi)(\nabla_\rho\nabla^\mu\varphi)  \right. \nonumber \\
&& \left. -6 (\nabla_\mu\varphi)(\nabla^\mu\nabla^\nu\varphi)(\nabla^\rho\varphi)G_{\nu\rho}\right],
\eq
where $\varphi$ is the Galileon field and $M^3\equiv M_{\rm Pl}H_0^2$ with $H_0$ being the present-day Hubble expansion rate. Note that the derivative couplings to the Ricci scalar $R$ and the Einstein tensor $G_{\mu\nu}$ in $\mathcal{L}_4$ and $\mathcal{L}_5$, respectively, break the shift symmetry.

{Besides the terms which appear in the Galileon Lagrangians, $\mathcal{L}_i$, we are also allowed to introduce a derivative coupling of the form $\mathcal{L}_{coupling} \sim G^{\mu\nu}\nabla_\mu\varphi\nabla_\nu\varphi$ with the equations remaining up to second-order \cite{Appleby:2011aa,  Sushkov:2009hk, Gubitosi:2011sg, deRham:2011by, VanAcoleyen:2011mj, Zumalacarregui:2012us, Amendola:1993uh}. In \cite{Appleby:2011aa} this term was considered in the context of the covariant Galileon model where it was shown that in the weak field limit, where the curvature is not too high, this coupling term in the Jordan frame can be cast in the form of an explicit coupling to matter fields in the Einstein frame.}

In the rest of the paper we will choose to work in the Jordan frame adding to Eq.~(\ref{Galileon action}) the Lagrangian density

\bq
\mathcal{L}_G = -c_G\frac{M_{\rm Pl}}{M^3}G^{\mu\nu}\nabla_\mu\varphi\nabla_\nu\varphi,
\eq
where $c_G$ is a dimensionless constant which determines the strength of the coupling. We will be interested in the cases where the acceleration is due only to the field kinetic terms and therefore we will set the potential term $c_1$ to zero. 

The modified Einstein equations and the Galileon equation of motion are obtained by varying the action, $S$, with respect to $g_{\mu\nu}$ and $\varphi$, respectively. Our derivation agrees with those present in the literature \citep{PhysRevD.79.084003,Appleby:2011aa} although we explicitly write the Riemann tensor in terms of the Ricci and Weyl tensors, whenever it leads to the cancellation of some terms and hence to a slight simplification of the final expressions. We show these equations in Appendix~\ref{Appendix A}.

\section{The perturbation equations}

\label{The perturbation equations}

\subsection{The Perturbed Equations in General Relativity}

In this section we derive the covariant and gauge invariant perturbation equations in Galileon gravity. This will be done in detail below but before that let us outline the main ingredients of $3+1$ decomposition and their application to general relativity for ease of later reference.

The main idea of $3+1$ decomposition is to make spacetime splits of physical quantities with respect to the 4-velocity $u^\mu$ of an observer. The projection tensor $h_{\mu\nu}$ is defined by $h_{\mu\nu} = g_{\mu\nu} - u_\mu u_\nu$ and can be used to obtain covariant tensors which live in 3-dimensional hyperspaces perpendicular to $u^\mu$. For example, the covariant spatial derivative $\hat{\nabla}$ of a tensor field $T^{\beta...\gamma}_{\sigma...\lambda}$ is defined as
\bq
\hat{\nabla}^\alpha T^{\beta\cdot\cdot\cdot\gamma}_{\sigma\cdot\cdot\cdot\lambda} \equiv h^{\alpha}_{\mu}h^{\beta}_{\nu}\cdot\cdot\cdot h^{\gamma}_{\kappa}h^{\rho}_{\sigma}\cdot\cdot\cdot h^{\eta}_{\lambda}\nabla^\mu T^{\nu\cdot\cdot\cdot\kappa}_{\rho\cdot\cdot\cdot\eta}.
\eq

The energy-momentum tensor and covariant derivative of the 4-velocity are decomposed, respectively, as 

\bq
\label{Tuv} T_{\mu\nu} &=& \pi_{\mu\nu} + 2q_{(\mu}u_{\nu)} + \rho u_\mu u_\nu - ph_{\mu\nu},\\
\nabla_\mu u_\nu &=& \sigma_{\mu\nu} + \varpi_{\mu\nu} + \frac{1}{3}\theta h_{\mu\nu} + u_\mu A_\nu,
\eq
where $\pi_{\mu\nu}$ is the projected symmetric and trace-free (PSTF) anisotropic stress, $q_\mu$ is the heat flux vector, $p$ is the isotropic pressure, $\rho$ is the energy density, $\sigma_{\mu\nu}$ the PSTF shear tensor, $\varpi_{\mu\nu} = \hat{\nabla}_{[\mu}u_{\nu]}$ the vorticity, $\theta = \nabla^\alpha u_\alpha = 3\dot{a}/a = 3H$ ($a$ is the mean expansion scale factor) the expansion scalar and $A_\mu = \dot{u}_\mu$; the overdot denotes a time derivative expressed as $\dot{\phi} = u^\alpha\nabla_\alpha\phi$, brackets mean antisymmetrization and parentheses symmetrization. The normalization is such that $u^\alpha u_\alpha = 1$. The quantities $\pi_{\mu\nu}$, $q_\mu$, $\rho$ and $p$ are referred to as dynamical quantities and $\sigma_{\mu\nu}$, $\varpi_{\mu\nu}$, $\theta$ and $A_\mu$ as kinematical quantities. Note that the dynamical quantities can be obtained from Eq.~(\ref{Tuv}) using the relations

\bq\label{Tuv projection}
\rho &=& T_{\mu\nu}u^\mu u^\nu, \nonumber \\
p &=& -\frac{1}{3}h^{\mu\nu}T_{\mu\nu}, \nonumber \\
q_\mu &=& h_\mu^\nu u^\rho T_{\nu\rho}, \nonumber \\
\pi_{\mu\nu} &=& h_\mu^\rho h_\nu^\tau T_{\rho\tau} + ph_{\mu\nu}.
\eq

Decomposing the Riemann tensor and making use of Einstein equations, we obtain, after linearization, five constraint equations \cite{Challinor:1998xk}:

\bq\label{c1}
0 &=& \hat{\nabla^\alpha}\left( \epsilon^{\mu\nu}_{\ \ \ \alpha\beta}u^\beta\varpi_{\mu\nu}\right), \\
\label{c2}
\kappa q_\mu &=& -\frac{2\hat{\nabla}_\mu\theta}{3} + \hat{\nabla}^\nu\sigma_{\mu\nu} + \hat{\nabla}^\nu\varpi_{\mu\nu},\\
\label{c3}
\mathcal{B}_{\mu\nu} &=& \left[ \hat{\nabla}^\alpha\sigma_{\beta(\mu} + \hat{\nabla}^\alpha\varpi_{\beta(\mu}\right]\epsilon_{\nu)\gamma\alpha}^{\ \ \ \ \beta} u^\gamma, \\
\label{c4}
\hat{\nabla}^\nu\mathcal{E}_{\mu\nu} &=& \frac{1}{2}\kappa\left[ \hat{\nabla}^\nu\pi_{\mu\nu} + \frac{2}{3}\theta q_\mu +  \frac{2}{3}\hat{\nabla}_\mu\rho\right], \\
\label{c5}
\hat{\nabla}^\nu\mathcal{B}_{\mu\nu} &=& \frac{1}{2}\kappa\left[ \hat{\nabla}_\alpha q_\beta + (\rho + p)\varpi_{\alpha\beta}\right]\epsilon_{\mu\nu}^{\ \ \alpha\beta}u^\nu;
\eq
and five propagation equations:
\bq\label{p1}
0 &=& \dot{\theta} + \frac{1}{3}\theta^2 - \hat{\nabla} \cdot A + \frac{\kappa}{2}(\rho + 3p), \\
\label{p2}
0 &=& \dot{\sigma}_{\mu\nu} + \frac{2}{3}\theta\sigma_{\mu\nu} - \hat{\nabla}_{\langle\mu}A_{\nu\rangle} + \mathcal{E}_{\mu\nu} + \frac{\kappa}{2}\pi_{\mu\nu}, \\
\label{p3}
0 &=&\dot{\varpi}_{\mu\nu} + \frac{2}{3}\theta\varpi_{\mu\nu} - \hat{\nabla}_{[\mu}A_{\nu]}, \\
\label{p4}
0 &=& \frac{\kappa}{2}\left[ \dot{\pi}_{\mu\nu} + \frac{1}{3}\theta\pi_{\mu\nu}\right] - \frac{\kappa}{2}\left[ (\rho + p)\sigma_{\mu\nu} + \hat{\nabla}_{\langle\mu}q_{\nu\rangle}\right] \nonumber \\
&& - \left[ \dot{\mathcal{E}}_{\mu\nu} + \theta\mathcal{E}_{\mu\nu} - \hat{\nabla}^\alpha\mathcal{B}_{\beta(\mu}\epsilon_{\nu)\gamma\alpha}^{\ \ \ \ \ \beta}u^\gamma\right], \\
\label{p5}
0 &=&\dot{\mathcal{B}}_{\mu\nu} + \theta\mathcal{B}_{\mu\nu} + \hat{\nabla}^\alpha\mathcal{E}_{\beta(\mu}\epsilon_{\nu)\gamma\alpha}^{\ \ \ \ \ \beta}u^\gamma \nonumber \\ 
&&+ \frac{\kappa}{2}\hat{\nabla}^\alpha\pi_{\beta(\mu}\epsilon_{\nu)\gamma\alpha}^{\ \ \ \ \ \beta}u^\gamma.
\eq
Here, $\epsilon_{\mu\nu\alpha\beta}$ is the covariant permutation tensor, $\mathcal{E}_{\mu\nu}$ and $\mathcal{B}_{\mu\nu}$ are, respectively, the electric and magnetic parts of the Weyl tensor $\mathcal{W}_{\mu\nu\alpha\beta}$, defined by $\mathcal{E}_{\mu\nu} = u^\alpha u^\beta\mathcal{W}_{\mu\alpha\nu\beta}$ and $\mathcal{B}_{\mu\nu} = -\frac{1}{2}u^\alpha u^\beta \epsilon_{\mu\alpha}^{\ \ \gamma\delta}\mathcal{W}_{\gamma\delta\nu\beta}$. The angle brackets mean taking the trace-free part of a quantity and $\hat{\nabla} \cdot v = \hat{\nabla}^\alpha v_\alpha$, where $v$ is an arbitrary vector.

Besides the above equations, it is useful to express the projected Ricci scalar $\hat{R}$ into the hypersurfaces orthogonal to $u^\mu$ as
\bq\label{spatial curvature}
\hat{R} = 2\kappa\rho - \frac{2}{3}\theta^2.
\eq
The spatial derivative of the projected Ricci scalar, $\eta_\mu \equiv a\hat{\nabla}_\mu\hat{R}/2$, is given as
\bq\label{derivative spatial curvature}
\eta_\mu = \kappa a \hat{\nabla}_\mu\rho - \frac{2a}{3}\theta\hat{\nabla}_\mu\theta,
\eq
and its propagation equation given by
\bq\label{propagation spatial curvature}
\dot{\eta}_\mu + \frac{2\theta}{3}\eta_\mu = -\frac{2a\theta}{3}\hat{\nabla}_\mu\hat{\nabla}\cdot A - a\kappa\hat{\nabla}_\mu\hat{\nabla}\cdot q.
\eq

Finally, there are the conservation equations for the energy-momentum tensor:
\bq\label{conservation1}
\dot{\rho} + (\rho + p)\theta + \hat{\nabla}\cdot q &=& 0, \\
\label{conservation2}
\dot{q}_\mu + \frac{4}{3}\theta q_{\mu} + (\rho+p)A_\mu - \hat{\nabla}_\mu p + \hat{\nabla}^\nu\pi_{\mu\nu} &=& 0.
\eq

In this paper we will always consider the case of a spatially-flat Universe and, as a result, {the spatial curvature vanishes at the background level. Thus, setting $\hat{R} = 0$ in Eq.~(\ref{spatial curvature})}, we obtain the first Friedmann equation

\bq\label{background1}
\frac{\theta^2}{3} = \kappa {\rho}.
\eq
Note that at the background level only the zeroth-order terms contribute to the equations. The second Friedmann equation and the energy-conservation equation are obtained by taking the zeroth-order parts of Eqs.~(\ref{p1}, \ref{conservation1}), as

\bq\label{background2}
\dot{\theta} + \frac{1}{3}\theta^2 + \frac{\kappa}{2}({\rho} + 3{p}) &=& 0, \\
\label{background3}
\dot{{\rho}} + ({\rho} + {p})\theta &=& 0.
\eq

In what follows, we will only consider scalar modes of perturbations, for which the vorticity, $\varpi_{\mu\nu}$, and the magnetic part of the Weyl tensor, $\mathcal{B}_{\mu\nu}$, are at most of second order \cite{Challinor:1998xk} and will be neglected from our first-order study.

\subsection{The Perturbation Quantities in Galileon Gravity}

In the effective energy-momentum tensor approach, the field equations Eqs.~(\ref{c1} - \ref{background3}) above preserve their forms, but the dynamical quantities $\rho$, $p$, $q_\mu$ and $\pi_{\mu\nu}$ should be replaced by the effective total ones $\rho^{tot} = \rho^f + \rho^G$, $p^{tot} = p^f + p^G$, $q^{tot}_\mu = q_\mu^f + q_\mu^G$ and $\pi^{tot}_{\mu\nu} = \pi_{\mu\nu}^f + \pi_{\mu\nu}^G$, in which the superscripts $^G$ and $^f$ identify the contributions from the Galileon field and the rest of the matter fluid (including cold dark matter, baryons, photons and neutrinos), respectively. From here on we shall drop the superscript $^{tot}$ for ease of notation.

Before using Eq.~(\ref{Tuv projection}) to calculate $\rho^G$, $p^G$, $q_\mu^G$ and $\pi_{\mu\nu}^G$ from the components of the Galileon energy-momentum tensor shown in Appendix \ref{Appendix A} , we need an explicit expression for the Ricci tensor $R_{\mu\nu}$ in terms of the kinematical quantities. For this let us expand the symmetric rank-2 tensor $R_{\mu\nu}$ in the following general way
\bq
R_{\mu\nu} &=& \Delta u_\mu u_\nu + \Xi h_{\mu\nu} + 2u_{(\mu}\Upsilon_{\nu)} + \Sigma_{\mu\nu},
\eq
in which $\Upsilon_\mu$ is a four-vector and $\Sigma_{\mu\nu}$ a PSTF rank-2 tensor, both of which live in the 3-dimensional hyperspace perpendicular to the observer's four-velocity ($u^\mu\Upsilon_\mu=u^\mu\Sigma_{\mu\nu}=0$). $\Delta$ and $\Xi$ are scalar quantities. Then, using the modified Einstein field equations
\bq
R_{\mu\nu} - \frac{1}{2}g_{\mu\nu}R = \kappa T_{\mu\nu}^{tot} = \kappa T_{\mu\nu}^f + \kappa T_{\mu\nu}^G,
\eq
one gets,

\bq
\Delta  &=& \frac{1}{2}\kappa\left( \rho + 3p\right) \nonumber \\
&=& - \left[ \dot{\theta} + \frac{1}{3}\theta^2 - \hat{\nabla}\cdot A\right], \\
\Xi &=& -\frac{1}{2}\kappa\left( \rho - p\right) \nonumber \\
&=& -\frac{1}{3}\left[ \dot{\theta} + \theta^2 +  \hat{R} - \hat{\nabla}\cdot A\right], \\
\Upsilon_\mu &=& \kappa q_\mu \nonumber \\
&=& -\frac{2\hat{\nabla}_\mu\theta}{3} + \hat{\nabla}^\nu\sigma_{\mu\nu}+ \hat{\nabla}^\nu\varpi_{\mu\nu},\\
\Sigma_{\mu\nu} &=& \kappa\pi_{\mu\nu} \nonumber \\
&=& -2 \left[ \dot{\sigma}_{\mu\nu} + \frac{2}{3}\theta\sigma_{\mu\nu} - \hat{\nabla}_{\langle\mu}A_{\nu\rangle} + \mathcal{E}_{\mu\nu}\right].
\eq
where we have used Eqs.~(\ref{c2}, \ref{p1}, \ref{p2}, \ref{spatial curvature}). Notice that the first lines are expressed in terms of total dynamical quantities and the second lines in terms of kinematical quantities.

With the above useful relations and after some tedious but straightforward calculations, the Galileon contribution to the energy-momentum tensor {up to first order in perturbed quantities} can be identified as
\begin{widetext}
\bq\label{perturbed1}
\rho^G &=& c_2\left[ \frac{1}{2}\dot{\varphi}^2\right] + \frac{c_3}{M^3} \left[ 2\dot{\varphi}^3\theta + 2\dot{\varphi}^2\hat{\Box}\varphi\right]  +\frac{c_4}{M^6}\left[ \frac{5}{2}\dot{\varphi}^4\theta^2 + 4\dot{\varphi}^3\theta\hat{\Box}\varphi + \frac{3}{4}\dot{\varphi}^4\hat{R}\right]     \nonumber \\
&& + \frac{c_5}{M^9}\left[\frac{7}{9}\dot{\varphi}^5\theta^3 + \frac{5}{3}\dot{\varphi}^4\theta^2\hat{\Box}\varphi +\frac{1}{2}\dot{\varphi}^5\theta\hat{R} \right] + \frac{M_{\rm Pl}}{M^3}c_G\left[ \dot{\varphi}^2\theta^2 + \frac{4}{3}\dot{\varphi}\theta\hat{\Box}\varphi + \frac{1}{2}\dot{\varphi}^2\hat{R}\right] {+ \rm{higher\ order\ terms}}, \\
\label{perturbed2}
p^G &=& c_2\left[ \frac{1}{2}\dot{\varphi}^2\right] + \frac{c_3}{M^3} \left[ -2\ddot{\varphi}\dot{\varphi}^2\right] \nonumber \\
&&  + \frac{c_4}{M^6}\left[ -4\ddot{\varphi}\dot{\varphi}^3\theta - \dot{\varphi}^4\dot{\theta} - \frac{1}{2}\dot{\varphi}^4\theta^2 - 4\ddot{\varphi}\dot{\varphi}^2\hat{\Box}\varphi - \frac{4}{9}\dot{\varphi}^3\theta\hat{\Box}\varphi + \dot{\varphi}^4\hat{\triangledown}\cdot A + \frac{1}{12}\dot{\varphi}^4\hat{R}\right] \nonumber \\
&& + \frac{c_5}{M^9}\left[-\frac{5}{3}\ddot{\varphi}\dot{\varphi}^4\theta^2 - \frac{2}{3}\dot{\varphi}^5\dot{\theta}\theta - \frac{2}{9}\dot{\varphi}^5\theta^3 - \frac{2}{9}\dot{\varphi}^4\theta^2\hat{\Box}\varphi - \frac{8}{3}\ddot{\varphi}\dot{\varphi}^3\theta\hat{\Box}\varphi - \frac{1}{2}\ddot{\varphi}\dot{\varphi}^4\hat{R} - \frac{2}{3}\dot{\varphi}^4\dot{\theta}\hat{\Box}\varphi + \frac{2}{3}\dot{\varphi}^5\theta\hat{\triangledown} \cdot A\right] \nonumber \\
&& + \frac{M_{\rm Pl}}{M^3}c_G\left[ - \frac{4}{3}\ddot{\varphi}\dot{\varphi}\theta - \frac{2}{3}\dot{\varphi}^2\dot{\theta} - \frac{1}{3}\dot{\varphi}^2\theta^2 + \frac{2}{3}\dot{\varphi}^2\hat{\triangledown}\cdot A - \frac{4}{3}\ddot{\varphi}\hat{\Box}\varphi - \frac{4}{9}\dot{\varphi}\theta\hat{\Box}\varphi + \frac{1}{6}\dot{\varphi}^2\hat{R}\right] { + \rm{higher\ order\ terms}}, \\
\label{perturbed3}
q_\mu^G &=& c_2\left[ \dot{\varphi}\hat{\triangledown}_\mu\varphi \right] + \frac{c_3}{M^3} \left[ 2\dot{\varphi}^2\theta\hat{\triangledown}_\mu\varphi - 2\dot{\varphi}^2\hat{\triangledown}_\mu\dot{\varphi}\right] \nonumber \\
&&  + \frac{c_4}{M^6}\left[ -4\dot{\varphi}^3\theta\hat{\triangledown}_\mu\dot{\varphi} + 2\dot{\varphi}^3\theta^2\hat{\triangledown}_\mu\varphi - \dot{\varphi}^4\hat{\triangledown}_\mu\theta + \frac{3}{2}\dot{\varphi}^4\hat{\triangledown}^\alpha\sigma_{\mu\alpha} + \frac{3}{2}\dot{\varphi}^4\hat{\triangledown}^\alpha{\varpi}_{\mu\alpha} \right] \nonumber \\
&& + \frac{c_5}{M^9}\left[ - \frac{5}{3}\dot{\varphi}^4\theta^2\hat{\triangledown}_\mu\dot{\varphi} + \frac{5}{9}\dot{\varphi}^4\theta^3\hat{\triangledown}_\mu\varphi - \frac{2}{3}\dot{\varphi}^5\theta\hat{\triangledown}_\mu\theta + \dot{\varphi}^5\theta\hat{\triangledown}^\alpha\sigma_{\mu\alpha} + \dot{\varphi}^5\theta\hat{\triangledown}^\alpha{\varpi}_{\mu\alpha} \right] \nonumber \\
&& + \frac{M_{\rm Pl}}{M^3}c_G\left[ -\frac{4}{3}\dot{\varphi}\theta\hat{\triangledown}_\mu\dot{\varphi} + \frac{2}{3}\dot{\varphi}\theta^2\hat{\triangledown}_\mu\varphi - \frac{2}{3}\dot{\varphi}^2\hat{\triangledown}_\mu\theta + \dot{\varphi}^2\hat{\triangledown}^\alpha\sigma_{\mu\alpha} + \dot{\varphi}^2\hat{\triangledown}^\alpha{\varpi}_{\mu\alpha}\right] { + \rm{higher\ order\ terms}}, \\
\label{perturbed4}
\pi_{\mu\nu}^G &=& \frac{c_4}{M^6}\left[  -\dot{\varphi}^4 \left( \dot{\sigma}_{\mu\nu} - \hat{\triangledown}_{\langle\mu}A_{\nu\rangle} - \mathcal{E}_{\mu\nu}\right) - \left( 6\ddot{\varphi}\dot{\varphi}^2 + \frac{2}{3}\dot{\varphi}^3\theta\right)\hat{\triangledown}_{\langle\mu}\hat{\triangledown}_{\nu\rangle}\varphi - \left(6\ddot{\varphi}\dot{\varphi}^3 + \frac{4}{3}\dot{\varphi}^4\theta\right)\sigma_{\mu\nu}\right] \nonumber \\
&& + \frac{c_5}{M^9}\left[ -\left(\dot{\varphi}^5\dot{\theta} +\dot{\varphi}^5\theta^2 + 6\ddot{\varphi}\dot{\varphi}^4\theta\right)\sigma_{\mu\nu} -  \left(\dot{\varphi}^5\theta + 3\ddot{\varphi}\dot{\varphi}^4\right)\dot{\sigma}_{\mu\nu} -  \left( 4\ddot{\varphi}\dot{\varphi}^3\theta + \dot{\varphi}^4\dot{\theta} + \frac{1}{3}\dot{\varphi}^4\theta^2 \right)\hat{\triangledown}_{\langle\mu}\hat{\triangledown}_{\nu\rangle}\varphi \right. \nonumber \\
&& \left. \ \ \ \ \ \ \ \ \ \ \ \ \ + \left(\dot{\varphi}^5\theta + 3\ddot{\varphi}\dot{\varphi}^4 \right)\hat{\triangledown}_{\langle\mu}A_{\nu\rangle} - 6\ddot{\varphi}\dot{\varphi}^4\mathcal{E}_{\mu\nu} \right] \nonumber \\
&& + \frac{M_{\rm Pl}}{M^3}c_G\left[ -\left( 2\ddot{\varphi}\dot{\varphi} + \frac{2}{3}\dot{\varphi}^2\theta\right)\sigma_{\mu\nu} - \left( \frac{2}{3}\dot{\varphi}\theta + 2\ddot{\varphi}\right) \hat{\triangledown}_{\langle\mu}\hat{\triangledown}_{\nu\rangle}\varphi + 2\dot{\varphi}^2\mathcal{E}_{\mu\nu}\right] {+ \rm{higher\ order\ terms}},
\eq
\end{widetext}
in which $\hat{\Box}\equiv\hat{\nabla}^\mu\hat{\nabla}_\mu$.

Following the same procure, the Galileon field equation of motion (see Appendix \ref{Appendix A}) is given by

\begin{widetext}
\begin{eqnarray}
\label{perturbed EoM}
0 &=& c_2\left[\ddot{\varphi} + \hat{\Box}\varphi + \dot{\varphi}\theta \right] + \frac{c_3}{M^3} \left[ 4\ddot{\varphi}\dot{\varphi}\theta  + \frac{8}{3}\dot{\varphi}\theta\hat{\Box}\varphi + 4\ddot{\varphi}\hat{\Box}\varphi + 2\dot{\varphi}^2\theta^2 + 2\dot{\varphi}^2\dot{\theta} - 2\dot{\varphi}^2 \hat{\nabla} \cdot A\right] \nonumber \\
&&  + \frac{c_4}{M^6}\left[  6\ddot{\varphi}\dot{\varphi}^2\theta^2 + 4\dot{\varphi}^3\dot{\theta}\theta + 2\dot{\varphi}^3\theta^3 + 8\ddot{\varphi}\dot{\varphi}\theta\hat{\Box}\varphi + \frac{26}{9}\dot{\varphi}^2\theta^2\hat{\Box}\varphi - 4\dot{\varphi}^3\theta\hat{\nabla}\cdot A + 4\dot{\varphi}^2\dot{\theta}\hat{\Box}\varphi + 3\ddot{\varphi}\dot{\varphi}^2\hat{R} + \frac{1}{3}\dot{\varphi}^3\theta\hat{R}\right] \nonumber \\
&& + \frac{c_5}{M^9}\left[ \frac{5}{9}\dot{\varphi}^4\theta^4   +\frac{20}{9}\ddot{\varphi}\dot{\varphi}^3\theta^3  +\frac{5}{3}\dot{\varphi}^4\dot{\theta}\theta^2 +\frac{8}{9}\dot{\varphi}^3\theta^3\hat{\Box}\varphi +\frac{1}{2}\dot{\varphi}^4\dot{\theta}\hat{R} \right. \nonumber \\
&&\ \ \ \ \ \ \ \ \ \ \ \ \ \ \ \ \ \  \left. +\frac{1}{6}\dot{\varphi}^4\theta^2\hat{R} - \frac{5}{3}\dot{\varphi}^4\theta^2\hat{\nabla} \cdot A  + 4\ddot{\varphi}\dot{\varphi}^2\theta^2\hat{\Box}\varphi + \frac{8}{3}\dot{\varphi}^3\dot{\theta}\theta\hat{\Box}\varphi + 2\ddot{\varphi}\dot{\varphi}^3\theta\hat{R} \right] \nonumber \\
&& + \frac{M_{\rm Pl}}{M^3}c_G\left[ \frac{2}{3}\ddot{\varphi}\theta^2 + \frac{4}{3}\dot{\theta}\hat{\Box}\varphi + \frac{2}{3}\theta^2\hat{\Box}\varphi + \frac{4}{3}\dot{\varphi}\dot{\theta}\theta + \frac{2}{3}\dot{\varphi}\theta^3 - \frac{4}{3}\dot{\varphi}\theta\hat{\nabla} \cdot A + \ddot{\varphi}\hat{R} + \frac{1}{3}\dot{\varphi}\theta\hat{R}\right] { + \rm{higher\ order\ terms}}.
\end{eqnarray}
\end{widetext}

As a consistency test, we checked that Eqs.~(\ref{perturbed1} - \ref{perturbed4}) satisfy the conservation Eqs.~(\ref{conservation1}, \ref{conservation2}).

\subsection{Perturbed Equations in $k$-space}

For the purpose of the numerical studies presented in this paper, we need to write the perturbed quantities derived in the last subsection in terms of $k$-space variables. This is achieved with the aid of the following harmonic definitions:

\bq
\hat{\nabla}_\mu\varphi &\equiv & \sum_k \frac{k}{a}\gamma Q_\mu^k, \ \ \ \ \hat{\nabla}_\mu\theta \equiv  \sum_k\frac{k^2}{a^2}\mathcal{Z}Q_\mu^k, \nonumber \\
A_\mu &\equiv & \sum_k\frac{k}{a}AQ_\mu^k,\ \ \ \ \ \hat{\nabla}_\mu\rho \equiv  \sum_k \frac{k}{a}\chi Q_\mu^k, \nonumber \\
{\pi}_{\mu\nu} &\equiv & \sum_k \Pi Q_{\mu\nu}^k, \ \ \ \  \ \sigma_{\mu\nu} \equiv  \sum_k \frac{k}{a}\sigma Q_{\mu\nu}^k, \nonumber \\
\eta_\mu &\equiv & \sum_k \frac{k^3}{a^2}\eta Q_\mu^k, \ \ \ \mathcal{E}_{\mu\nu} \equiv  - \sum_k \frac{k^2}{a^2}\phi Q_{\mu\nu}^k,
\eq
in which $Q^k$ is the eigenfunction of the comoving spatial Laplacian $a^2\hat{\Box}$ satisfying

\bq
\hat{\Box}Q^k = \frac{k^2}{a^2}Q^k,
\eq
and $Q_\mu^k$ and $Q_{\mu\nu}^k$ are given by $Q_\mu^k = \frac{a}{k}\hat{\nabla}_\mu Q^k$ and by $Q_{\mu\nu}^k = \frac{a}{k}\hat{\nabla}_{\langle\mu}Q_{\nu\rangle}$, respectively.

In terms of these harmonic expansion variables, Eqs.~(\ref{c2}, \ref{c4}, \ref{p2}, \ref{p4}, \ref{derivative spatial curvature}, \ref{propagation spatial curvature}) can be rewritten as 

\bq
\frac{2}{3}k^2(\sigma - \mathcal{Z}) &=& \kappa qa^2, \\
k^3\phi &=& -\frac{1}{2}\kappa a^2 \left[ k(\Pi + \chi) + 3\mathcal{H}q \right], \\
k(\sigma\prime + \mathcal{H}\sigma) &=& k^2(\phi + A) - \frac{1}{2}\kappa a^2\Pi, \\
k^2(\phi^\prime + \mathcal{H}\phi) &=& \frac{1}{2}\kappa a^2 \left[ k(\rho + p)\sigma + kq - \Pi^\prime - \mathcal{H}\Pi \right], \\
k^2\eta &=& \kappa \chi a^2 - 2k\mathcal{H}\mathcal{Z}, \\
k\eta\prime &=& -\kappa q a^2 - 2k\mathcal{H}A,
\eq
respectively, where $\mathcal{H} = a'/a$ and a prime denotes a derivative with respect to conformal time $\tau$ ($ad\tau = dt$, with $t$ the physical time). From Eqs.~(\ref{perturbed1}, \ref{perturbed3}, \ref{perturbed4})  one obtains the $k$-space variables $\chi^G$, $q^G$ and $\Pi^G$

\begin{widetext}
\bq
\chi^G &=& c_2\frac{1}{a^2}\left( \varphi'\gamma' + \varphi'^2A\right) + \frac{c_3}{M^3}\frac{1}{a^4} \left( \left[ 18\varphi'^2\mathcal{H}\gamma' + 18\varphi'^3\mathcal{H}A\right] + k\left[ 2\varphi'^3\mathcal{Z}\right] + k^2\left[ 2\varphi'^2\gamma\right]\right) \nonumber \\
&& + \frac{c_4}{M^6}\frac{1}{a^6} \left( \left[ 90\varphi'^3\mathcal{H}^2\gamma' + 90\varphi'^4\mathcal{H}^2A \right] + k\left[ 15\varphi'^4\mathcal{H}\mathcal{Z} \right] + k^2\left[ 12\varphi'^3\mathcal{H}\gamma + \frac{3}{2}\varphi'^4\eta\right]\right) \nonumber \\
&& + \frac{c_5}{M^9}\frac{1}{a^8}\left( \left[ 105\varphi'^4\mathcal{H}^3\gamma' + 105\varphi'^5\mathcal{H}^3A\right] + k\left[ 21\varphi'^5\mathcal{H}^2\mathcal{Z}\right] + k^2\left[ 15\varphi'^4\mathcal{H}^2\gamma + 3\varphi'^5\mathcal{H}\eta\right]\right) \nonumber \\
&& + \frac{M_{\rm Pl}}{M^3}c_G\frac{1}{a^4}\left( \left[ 18\varphi'\mathcal{H}^2\gamma' + 18\varphi'^2\mathcal{H}^2A\right] + k\left[ 6\varphi'^2\mathcal{H}\mathcal{Z}\right] + k^2\left[ 4\varphi'\mathcal{H}\gamma + \varphi'^2\eta\right]\right), \\
q^G &=& c_2 \frac{k}{a^2}\left(\varphi'\gamma\right) + \frac{c_3}{M^3}\frac{k}{a^4} \left( 6\varphi'^2\mathcal{H}\gamma - 2\varphi'^2\gamma' - 2\varphi'^3A \right) \nonumber \\
&& + \frac{c_4}{M^6}\frac{1}{a^6} \left( k\left[ -12\varphi'^3\mathcal{H}\gamma' - 12\varphi'^4\mathcal{H}A + 18\varphi'^3\mathcal{H}^2\gamma\right] + k^2\left[ \varphi'^4\sigma - \varphi'^4\mathcal{Z}\right]\right) \nonumber \\
&& + \frac{c_5}{M^9}\frac{1}{a^8}\left( k\left[ -15\varphi'^4\mathcal{H}^2\gamma' - 15\varphi'^5\mathcal{H}^2A + 15\varphi'^4\mathcal{H}^3\gamma\right] + 2k^2\left[ -\varphi'^5\mathcal{H}\mathcal{Z} + \varphi'^5\mathcal{H}\sigma\right]\right) \nonumber \\
&& + \frac{M_{\rm Pl}}{M^3}c_G\frac{1}{a^4}\left( k\left[-4\varphi' \mathcal{H}\gamma'  - 4\varphi'^2\mathcal{H}A + 6\varphi' \mathcal{H}^2\gamma \right] + \frac{2}{3}k^2\left[ \varphi'^2\sigma - \varphi'^2\mathcal{Z}\right]\right), \\
\Pi^G &=& \frac{c_4}{M^6}\frac{1}{a^6} \left( k\left[ -\varphi'^4\sigma' + 3\varphi'^4\mathcal{H}\sigma - 6\varphi''\varphi'^3\sigma\right] + k^2\left[ 4\varphi'^3\mathcal{H}\gamma - 6\varphi''\varphi'^2\gamma + \varphi'^4A - \varphi'^4\phi\right]\right) \nonumber \\
&& + \frac{c_5}{M^9}\frac{1}{a^8}\left( k\left[ -3\varphi'^5\mathcal{H}'\sigma + 12\varphi'^5\mathcal{H}^2\sigma - 15\varphi''\varphi'^4\mathcal{H}\sigma - 3\varphi''\varphi'^4\sigma'\right] \right. \nonumber \\
&& + \left. k^2\left[ -12\varphi''\varphi'^3\mathcal{H}\gamma + 12\varphi'^4\mathcal{H}^2\gamma - 3\varphi'^4\mathcal{H}'\gamma + 3\varphi''\varphi'^4A + 6\varphi''\varphi'^4\phi - 6\varphi'^5\mathcal{H}\phi \right]\right) \nonumber \\
&& + \frac{M_{\rm Pl}}{M^3}c_G\frac{1}{a^4}\left( k\left[ -2\varphi''\varphi'\sigma\right] -2 k^2\left[ \varphi''\gamma + \varphi'^2\phi\right]\right).
\eq
\end{widetext}
Note that the spatial derivative of the isotropic pressure $p$ in $k$-space is not needed in the {\tt CAMB} code, which is why we do not write it here. Finally, in $k$-space, the perturbed Galileon field equation of motion, Eq.~(\ref{perturbed EoM}), reads

\begin{widetext}
\bq
0 &=& \frac{c_2}{a^3} \left( k\left[ \gamma'' + 2\gamma'\mathcal{H} + \varphi' A' + \varphi'\mathcal{H}A + 2\varphi'' A\right] + k^2\varphi'\mathcal{Z} + k^3\gamma\right) \nonumber \\
&& +\frac{c_3}{M^3}\frac{1}{a^5} \left( k\left[ 12\gamma''\varphi'\mathcal{H} + 12\varphi'^2\mathcal{H}A' - 18\varphi'^2\mathcal{H}^2A + 36\varphi''\varphi'\mathcal{H}A + 12\varphi''\mathcal{H}\gamma' + 12\varphi'\mathcal{H}'\gamma' + 18\varphi'^2\mathcal{H}' A\right] \right. \nonumber \\
&& \ \ \ \ \ \ \ \ \ \ \ \ \ \ \ \ \ \ \ \ \ \left. k^2\left[6\varphi'^2\mathcal{H}\mathcal{Z} + 2\varphi'^2\mathcal{Z}' + 4\varphi''\varphi'\mathcal{Z}\right] + k^3\left[ 4\varphi'\mathcal{H}\gamma - 2\varphi'^2A + 4\varphi''\gamma\right]\right) \nonumber \\
&& +\frac{c_4}{M^6}\frac{1}{a^7} \left( k\left[ 54\varphi'^2\mathcal{H}^2\gamma'' - 108\varphi'^2\mathcal{H}^3\gamma' + 54\varphi'^3\mathcal{H}^2A' - 198\varphi'^3\mathcal{H}^3A + 216\varphi'' \varphi'^2\mathcal{H}^2A + 108\varphi''\varphi'\mathcal{H}^2\gamma'\right. \right. \nonumber \\
&& \ \ \ \ \ \ \ \ \ \ \ \ \ \ \ \ \ \ \ \ \ \ \ \ \left. \left. + 108\varphi'^2\mathcal{H}\mathcal{H}'\gamma' + 144\varphi'^3\mathcal{H}\mathcal{H}' A\right] + k^2\left[ -6\varphi'^3\mathcal{H}^2\mathcal{Z} + 36\varphi''\varphi'^2\mathcal{H}\mathcal{Z} + 12\varphi'^3\mathcal{H}'\mathcal{Z} + 12\varphi'^3\mathcal{H}\mathcal{Z}'\right] \right. \nonumber \\
&& \ \ \ \ \ \ \ \ \ \ \ \ \ \ \ \ \ \ \ \ \ \left. k^3 \left[ -10\varphi'^2\mathcal{H}^2\gamma - 12\varphi'^3\mathcal{H}A - 4\varphi'^3\mathcal{H}\eta + 24\varphi''\varphi'\mathcal{H}\gamma + 12\varphi'^2\mathcal{H}'\gamma + 6\varphi''\varphi'^2\eta\right]\right) \nonumber \\
&& +\frac{c_5}{M^9}\frac{1}{a^9} \left( k\left[ -240\varphi'^3\mathcal{H}^4\gamma' - 345\varphi'^4\mathcal{H}^4A + 60\varphi'^3\mathcal{H}^3\gamma'' + 60\varphi'^4\mathcal{H}^3A' + 300\varphi''\varphi'^3\mathcal{H}^3A  +180\varphi''\varphi'^2\mathcal{H}^3\gamma'\right. \right. \nonumber \\
&& \ \ \ \ \ \ \ \ \ \ \ \ \ \ \ \ \ \ \ \ \ \ \ \ \ \left. \left.  + 180\varphi'^3\mathcal{H}^2\mathcal{H}'\gamma' + 225\varphi'^4\mathcal{H}^2\mathcal{H}' A\right] + k^2\left[ -45\varphi'^4\mathcal{H}^3\mathcal{Z} + 60\varphi''\varphi'^3\mathcal{H}^2\mathcal{Z} + 15\varphi'^4\mathcal{H}^2\mathcal{Z}' + 30\varphi'^4\mathcal{H}\mathcal{H}'\mathcal{Z} \right] \right. \nonumber \\
&& \ \ \ \ \ \ \ \ \ \ \ \ \ \ \ \ \ \ \ \ \ \left. k^3\left[ -36\varphi'^3\mathcal{H}^3\gamma - 12\varphi'^4\mathcal{H}^2\eta - 15\varphi'^4\mathcal{H}^2A + 3\varphi'^4\mathcal{H}'\eta + 36\varphi''\varphi'^2\mathcal{H}^2\gamma + 24\varphi'^3\mathcal{H}\mathcal{H}'\gamma + 12\varphi''\varphi'^3\mathcal{H}\eta\right] \right) \nonumber \\
&&+\frac{M_{\rm Pl}}{M^3}\frac{c_G}{a^5} \left(k\left[ 6\mathcal{H}^2\gamma'' + 6\varphi'\mathcal{H}^2A' - 18\varphi'\mathcal{H}^3A + 12\varphi''\mathcal{H}^2A + 12\mathcal{H}\mathcal{H}'\gamma' + 24\varphi'\mathcal{H}\mathcal{H}' A\right] \right. \nonumber \\
&& \ \ \ \ \ \ \ \ \ \ \ \ \ \ \ \ \ \ \ \ \ \left. k^2\left[ 6\varphi'\mathcal{H}^2\mathcal{Z} + 4\varphi''\mathcal{H}\mathcal{Z} + 4\varphi'\mathcal{H}\mathcal{Z}' + 4\varphi'\mathcal{H}'\mathcal{Z}\right] + k^3\left[ 2\mathcal{H}^2\gamma - 4\varphi'\mathcal{H}A + 4\mathcal{H}'\gamma + 2\varphi''\eta\right]\right).
\eq
\end{widetext}

As another consistency test, we have checked that the conservation Eqs.~(\ref{conservation1}, \ref{conservation2}) in $k$-space,

\bq
\label{conservationk1}
\chi' + (k\mathcal{Z} - 3\mathcal{H}A)(\rho + p) + 3\mathcal{H}(\chi + \chi^p) + kq &=& 0, \\
\label{conservationk2}
q' + 4\mathcal{H}q + (\rho + p)kA - k\chi^p + \frac{2}{3}k\Pi &=& 0,
\eq
are satisfied by the $k$-space perturbed expressions derived above.

\subsubsection{Synchronous and Newtonian Gauge Equations}

Here, we present the recipe to write the CGI perturbation equations in the synchronous and in the newtonian gauge \cite{Ma:1994dv}.

The perturbed Friedmann-Robertson-Walker line element in the synchronous gauge is written as 

\bq
{ds^2}_S = a^2(\tau) \left[ d\tau^2 - (\delta_{ij} + h_{ij}^S)dx^idx^j\right].
\eq
Latin indices run over $1$, $2$ and $3$, $\delta_{ij}$ is the delta function and the spatial perturbed metric $h_{ij}^S \equiv h_{ij}^S(\bold{x}, \tau)$ is given by

\bq
h_{ij}^S = \int d^3k\ e^{i\bold{k}\bold{x}} && \left[ \hat{k_i}\hat{k_j}h^S(\bold{k}, \tau) \right. \nonumber \\
&& \left. + 6\left( \hat{k_i}\hat{k_j} - \frac{1}{3}\delta_{ij}\right)\eta^S(\bold{k}, \tau)  \right],
\eq
where a superscript $'S'$ denotes quantities in the synchronous gauge, $\bold{x}$ is the spatial position vector and $\hat{k} = \bold{k}/k$ is the unit vector mode in the $\bold{k}$-direction. The CGI and the synchronous gauge quantities are related by means of the following relations

\bq
\phi &=& \frac{1}{4k^2}\left[ 6\eta''^S + h''^S\right] - \frac{1}{4}\eta^S, \nonumber \\
A &=& 0, \nonumber \\
\eta &=& -2\eta^S, \nonumber \\ 
\mathcal{Z} &=& \frac{h'^{S}}{2k}, \nonumber  \\
\sigma &=& \frac{1}{2k}\left( 6\eta'^S + h'^S\right).
\eq

The line element in the Newtonian (also known as longitudinal) gauge is diagonal, described by two scalar potentials $\Psi$ and $\Phi$, and reads

\bq
ds^2_N = a^2(\tau) \left[ (1 + 2\Psi)d\tau^2 - (1-2\Phi)dx^idx_i\right].
\eq
Written in this way, the perturbed line element is only applicable to the study of the scalar modes of the metric perturbations. The two potentials are related to the Weyl potential $\phi$ as

\bq
\Psi &=& \phi - \frac{1}{2}\left(\frac{a}{k}\right)^2\kappa\Pi, \nonumber \\
\Phi &=& -\phi - \frac{1}{2}\left(\frac{a}{k}\right)^2\kappa\Pi,
\eq
and the other CGI quantities are given by

\bq
A &=& -\Psi, \nonumber \\
\eta &=& 2\Phi, \nonumber \\
\mathcal{Z} &=& \frac{3}{k} \left( \Phi' - \Psi\mathcal{H}\right), \nonumber \\
\sigma &=& 0.
\eq

We do not present the full perturbed field equations in the synchronous and Newtonian gauges because they are not used in our modified {\tt CAMB} code. However, note that {\tt CAMB} works in the cold-dark-matter frame where $A=0$, which is equivalent to the synchronous gauge written in a slightly different formalism.

\section{Results}\label{Results}

 In this section we present and discuss our results, which were obtained using a version of the {\tt CAMB} code \cite{camb_notes} suitably modified by us to follow Galileon gravity models.

\subsection{Background}

\begin{table}
\caption{The model parameters for the Galileon models studied in this paper. The $c_2$ parameter is tuned to yield the required amount of dark energy today and its exact value depends on the choice of the initial Galileon energy density $\rho_{\varphi,i}$.}
\begin{tabular}{@{}lcccccc}
\hline\hline
Models & \ \ \ \ \ \ \ \ \ \ \ \ $c_3$ & \ \ \ \ \ \ \ \ \ \ \ \ \ \ \ $c_4$ & \ \ \ \ \ \ \ \ \ \ \ \ \ $c_5$ & \ \ \ \ \ \ \ \ \ \ \ \ $c_G$ \\
\hline
Galileon 1 & \ \ \ \ \ \ \ \ \ \ \ \ 12.8 &\ \ \ \ \ \ \ \ \ \ \ \  $-1.7$ & \ \ \ \ \ \ \ \ \ \ \ \ 1.0 & \ \ \ \ \ \ \ \ \ \ \ \ $0$  \\
Galileon 2 & \ \ \ \ \ \ \ \ \ \ \ \ 6.239 & \ \ \ \ \ \ \ \ \ \ \ \ $-2.159$ &\ \ \ \ \ \ \ \ \ \ \ \  1.0 & \ \ \ \ \ \ \ \ \ \ \ \ $0$  \\
Galileon 3 & \ \ \ \ \ \ \ \ \ \ \ \ 5.73 & \ \ \ \ \ \ \ \ \ \ \ \ $-1.2$ & \ \ \ \ \ \ \ \ \ \ \ \ 1.0 &\ \ \ \ \ \ \ \ \ \ \ \  $0$ \\
Galileon 4 & \ \ \ \ \ \ \ \ \ \ \ \ 5.73 &\ \ \ \ \ \ \ \ \ \ \ \  $-1.2$ & \ \ \ \ \ \ \ \ \ \ \ \ 1.0 &\ \ \ \ \ \ \ \ \ \ \ \  $-0.4$ \\
\hline
\end{tabular}
\label{Models}
\end{table}

\begin{table}
\caption{The values of the parameter $c_2$ and of the age of the Universe for all the initial conditions used in this paper. The age for $\Lambda$CDM is 13.738 Gyr.}
\begin{tabular}{@{}lcccccc}
\hline\hline
$\rho_{\varphi, i} / \rho_{m,i}$ & \ \ \ \ \ \ \ \ \ \ \ \ $c_2$ & \ \ \ \ \ \ \ \ \ \ \ \ \ \ \ Age (Gyr) & \\
\hline
& \ \ \ \ \ \ \ \ \ \ \ \ \ \ \ Galileon 1   & & \\ 
\hline
\ \ \ \ $10^{-4}$ & \ \ \ \ \ \ \ \ \ \ \ $-27.00$ &\ \ \ \ \ \ \ \ \ \ \ \ \ \ \ \ \ \  $13.978$ & \\
\ \ \ \ $10^{-5}$ & \ \ \ \ \ \ \ \ \ \ \ $-27.49$ &\ \ \ \ \ \ \ \ \ \ \ \ \ \ \ \ \ \  $14.317$ & \\
\ \ \ \ $10^{-6}$ & \ \ \ \ \ \ \ \ \ \ \ $-27.56$ &\ \ \ \ \ \ \ \ \ \ \ \ \ \ \ \ \ \  $14.366$ & \\
\ \ \ \ $10^{-7}$ & \ \ \ \ \ \ \ \ \ \ \ $-27.58$ &\ \ \ \ \ \ \ \ \ \ \ \ \ \ \ \ \ \  $14.374$ & \\
\ \ \ \ $10^{-8}$ & \ \ \ \ \ \ \ \ \ \ \ $-27.59$ &\ \ \ \ \ \ \ \ \ \ \ \ \ \ \ \ \ \  $14.375$ & \\
\hline
& \ \ \ \ \ \ \ \ \ \ \ \ \ \ \ Galileon 2   & & \\
\hline
\ \ \ \ $10^{-4}$ & \ \ \ \ \ \ \ \ \ \ \ $-12.600$ &\ \ \ \ \ \ \ \ \ \ \ \ \ \ \ \ \ \  $13.614$ & \\
\ \ \ \ $10^{-5}$ & \ \ \ \ \ \ \ \ \ \ \ $-12.846$ &\ \ \ \ \ \ \ \ \ \ \ \ \ \ \ \ \ \  $14.256$ & \\
\ \ \ \ $5\times10^{-6}$ & \ \ \ \ \ \ \ \ \ \ \ $-12.857$ &\ \ \ \ \ \ \ \ \ \ \ \ \ \ \ \ \ \  $14.286$ & \\
\ \ \ \ $10^{-6}$ & \ \ \ \ \ \ \ \ \ \ \ $-12.885$ &\ \ \ \ \ \ \ \ \ \ \ \ \ \ \ \ \ \  $14.357$ & \\
\ \ \ \ $10^{-7}$ & \ \ \ \ \ \ \ \ \ \ \ $-12.891$ &\ \ \ \ \ \ \ \ \ \ \ \ \ \ \ \ \ \  $14.372$ & \\
\ \ \ \ $10^{-8}$ & \ \ \ \ \ \ \ \ \ \ \ $-12.892$ &\ \ \ \ \ \ \ \ \ \ \ \ \ \ \ \ \ \  $14.375$ & \\
\hline
& \ \ \ \ \ \ \ \ \ \ \ \ \ \ \ Galileon 3   & & \\
\hline
\ \ \ \ $10^{-4}$ & \ \ \ \ \ \ \ \ \ \ \ $-14.760$ &\ \ \ \ \ \ \ \ \ \ \ \ \ \ \ \ \ \  $13.854$ & \\
\ \ \ \ $10^{-5}$ & \ \ \ \ \ \ \ \ \ \ \ $-15.122$ &\ \ \ \ \ \ \ \ \ \ \ \ \ \ \ \ \ \  $14.296$ & \\
\ \ \ \ $10^{-6}$ & \ \ \ \ \ \ \ \ \ \ \ $-15.179$ &\ \ \ \ \ \ \ \ \ \ \ \ \ \ \ \ \ \  $14.363$ & \\
\ \ \ \ $10^{-7}$ & \ \ \ \ \ \ \ \ \ \ \ $-15.188$ &\ \ \ \ \ \ \ \ \ \ \ \ \ \ \ \ \ \  $14.373$ & \\
\ \ \ \ $10^{-8}$ & \ \ \ \ \ \ \ \ \ \ \ $-15.189$ &\ \ \ \ \ \ \ \ \ \ \ \ \ \ \ \ \ \  $14.375$ & \\
\hline
& \ \ \ \ \ \ \ \ \ \ \ \ \ \ \ Galileon 4  & & \\
\hline
\ \ \ \ $10^{-4}$ & \ \ \ \ \ \ \ \ \ \ \ $-14.186$ &\ \ \ \ \ \ \ \ \ \ \ \ \ \ \ \ \ \  $13.833$ & \\
\ \ \ \ $10^{-5}$ & \ \ \ \ \ \ \ \ \ \ \ $-14.519$ &\ \ \ \ \ \ \ \ \ \ \ \ \ \ \ \ \ \  $14.285$ & \\
\ \ \ \ $5\times10^{-6}$ & \ \ \ \ \ \ \ \ \ \ \ $-14.539$ &\ \ \ \ \ \ \ \ \ \ \ \ \ \ \ \ \ \  $14.312$ & \\
\hline
\end{tabular}
\label{initial conditions}
\end{table}

\begin{figure*}
	\centering
	\includegraphics[scale=0.3]{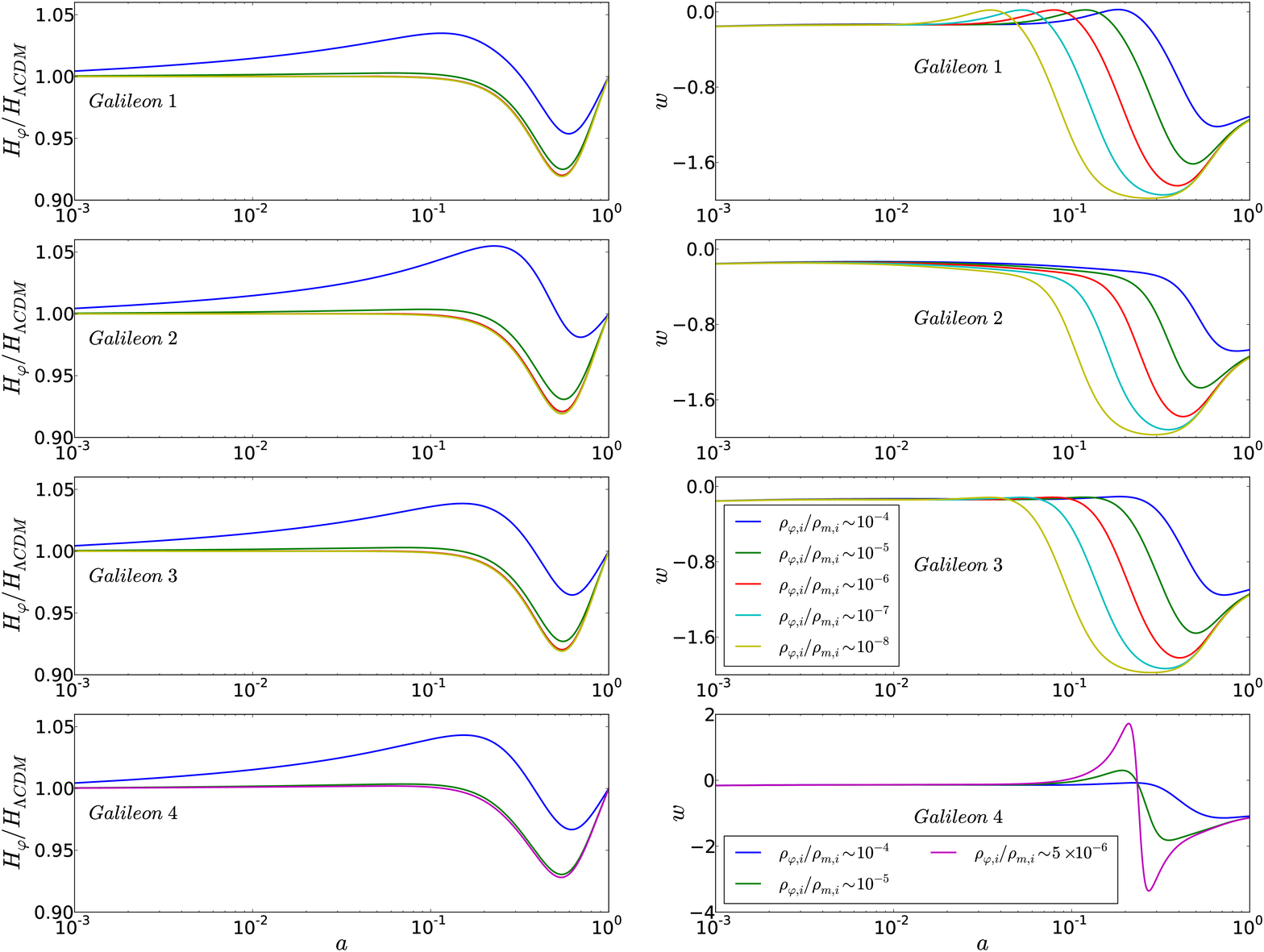}
	\caption{(Color online) Evolution of the ratio of the Hubble expansion rates of the Galileon and $\Lambda$CDM models, $H / H_{\Lambda CDM}$ ($H = \theta/3$), and of the Galileon field equation of state parameter $w$. The evolutions are shown for the four models of Table~\ref{Models} for different initial conditions. In the Galileon 1, Galileon 2 and Galileon 3 panels, on the left-hand side from top to bottom and on the right-hand side from right to left, the lines correspond, respectively, to $\rho_{\varphi, i}/\rho_{m,i} = \{10^{-4},10^{-5},10^{-6},10^{-7},10^{-8}\}$. The same for the Galileon 4 panels but for $\rho_{\varphi, i}/\rho_{m,i} = \{10^{-4},10^{-5},5\times10^{-6}\}$.}
\label{background}\end{figure*}

We compute the evolution of the cosmological background using the Friedmann equation, Eq.~(\ref{background2}), and the background Galileon equation of motion given by taking the zeroth-order terms of Eq.~(\ref{perturbed EoM}):

\bq\label{background EoM}
0 &=& c_2\left[\ddot{\varphi} + \dot{\varphi}\theta \right] + \frac{c_3}{M^3} \left[ 4\ddot{\varphi}\dot{\varphi}\theta + 2\dot{\varphi}^2\theta^2 + 2\dot{\varphi}^2\dot{\theta}\right] \nonumber \\
&&  + \frac{c_4}{M^6}\left[  6\ddot{\varphi}\dot{\varphi}^2\theta^2 + 4\dot{\varphi}^3\dot{\theta}\theta + 2\dot{\varphi}^3\theta^3 \right] \nonumber \\
&& + \frac{c_5}{M^9}\left[ \frac{5}{9}\dot{\varphi}^4\theta^4   +\frac{20}{9}\ddot{\varphi}\dot{\varphi}^3\theta^3  +\frac{5}{3}\dot{\varphi}^4\dot{\theta}\theta^2 \right] \nonumber \\
&& + \frac{M_{\rm Pl}}{M^3}c_G\left[ \frac{2}{3}\ddot{\varphi}\theta^2 + \frac{4}{3}\dot{\varphi}\dot{\theta}\theta + \frac{2}{3}\dot{\varphi}\theta^3\right].
\eq
The value of the Galilean background energy density $\bar{\rho}_{\varphi,i}$ at the starting redshift, which we take to be $z_i = 10^{6}$, is determined through the zeroth-order part of Eq.~(\ref{perturbed1}),

\bq\label{density-background}
& \bar{\rho}_\varphi = c_2\left[ \frac{1}{2}\dot{\varphi}^2\right] + \frac{c_3}{M^3} \left[ 2\dot{\varphi}^3\theta\right]  +\frac{c_4}{M^6}\left[ \frac{5}{2}\dot{\varphi}^4\theta^2\right]     \nonumber \\
& + \frac{c_5}{M^9}\left[\frac{7}{9}\dot{\varphi}^5\theta^3\right] + \frac{M_{\rm Pl}}{M^3}c_G\left[ \dot{\varphi}^2\theta^2\right],
\eq
by the initial values of the field time derivative $\dot{\varphi}_i$ and the expansion rate $\theta_i$, the latter being given by the fixed matter and radiation components via Eq.~(\ref{background1}) (the Galileon background energy density is negligible at early times). We specify $\theta_i$ using $\Omega_{m0} = 0.265$ and $\Omega_{r0} \approx 8\times10^{-5}$ for the present day values of the fractional energy density of matter and radiation, respectively \cite{Larson:2010gs, 2009MNRAS.400.1643S}. Since we are assuming a spatially flat Universe we need the evolution of the Galileon field to be such that $\Omega_{\varphi0} \approx 1 - \Omega_{m0} \approx 0.735$. This can be done by choosing appropriately the value of the $c_2$ parameter by a trial and error approach. As a consistency test, we have checked that Eqs.~(\ref{background1}, \ref{background3}) are satisfied by the numerical solution we obtain from {\tt CAMB}. Moreover, we have also checked that the background expansion solution from $\tt CAMB$ agrees very well with those in the literature \cite{Gannouji:2010au, PhysRevD.80.024037, DeFelice:2010pv, Nesseris:2010pc, Appleby:2011aa, PhysRevD.82.103015} and from an independent code written in {\tt Python} by us.

In this paper we focus on four different sets of Galileon parameters which we list in Table \ref{Models}. In \cite{Appleby:2011aa} (to which we refer the reader for further details on the background evolution of these models) it was shown that these choices of parameters are free of ghost and Laplace instabilities for initial conditions with $\rho_{\varphi,i} / \rho_{m,i} \sim 10^{-5}$. Here we shall use this and other choices of initial conditions which have not shown any theoretical instabilities {of the scalar perturbations throughout the entire expansion history} yielding therefore viable cosmological evolutions. In Table \ref{initial conditions} we list all these initial conditions with the values of the $c_2$ parameter and age of the Universe. {The initial conditions were chosen to span over a wide range of different behaviors of the Galileon model. It would be interesting to investigate the theoretical motivation and naturalness of these initial conditions although such an investigation is beyond the scope of the present work (see e.g. \cite{Pujolas:2011he, DeFelice:2011hq})}.

Figure \ref{background} shows the time evolution of the ratio of the Hubble expansion rates, $H = \theta/3$, of the Galileon and $\Lambda$CDM models and of the Galileon field  equation-of-state parameter, $w = \bar{p}_\varphi/\bar{\rho}_\varphi$, where

\bq\label{pressure-background}
& \bar{p}_\varphi = c_2\left[ \frac{1}{2}\dot{\varphi}^2\right] + \frac{c_3}{M^3} \left[ -2\ddot{\varphi}\dot{\varphi}^2\right] \nonumber \\
&  + \frac{c_4}{M^6}\left[ -4\ddot{\varphi}\dot{\varphi}^3\theta - \dot{\varphi}^4\dot{\theta} - \frac{1}{2}\dot{\varphi}^4\theta^2\right] \nonumber \\
& + \frac{c_5}{M^9}\left[-\frac{5}{3}\ddot{\varphi}\dot{\varphi}^4\theta^2 - \frac{2}{3}\dot{\varphi}^5\dot{\theta}\theta - \frac{2}{9}\dot{\varphi}^5\theta^3\right] \nonumber \\
& + \frac{M_{\rm Pl}}{M^3}c_G\left[ - \frac{4}{3}\ddot{\varphi}\dot{\varphi}\theta - \frac{2}{3}\dot{\varphi}^2\dot{\theta} - \frac{1}{3}\dot{\varphi}^2\theta^2\right],
\eq
is the background pressure (the zeroth-order part of Eq.~(\ref{perturbed2}).

Figure \ref{background} shows that, depending on the initial condition, the expansion rate can be faster or slower than in $\Lambda$CDM for different times during the evolution. {Another noteworthy aspect of the background evolution is the possibility of having ghost-free phantom dynamics, $w < -1$ \cite{Deffayet:2010qz, Pujolas:2011he, Sawicki:2012re}}. The initial values of $\rho_{\varphi,i}$ can have a great impact on the evolution of $w$: the lower $\rho_{\varphi,i}$ the more negative the values of $w$ will be. The reason is that lower values of $\rho_\varphi$ in the past will force the energy density to grow more drastically ($w < -1$) closer to today when the field starts to be driven towards the de Sitter attractor evolution \cite{DeFelice:2010pv,Appleby:2011aa, Appleby:2012ba} (see \cite{Nesseris:2010pc, PhysRevD.82.103015} for expansion history observational constraints). However, for $\rho_{\varphi,i}\lesssim10^{-5}$, the strong dependence of $w$ on the initial conditions does not propagate into the expansion rate which is only sensitive to changes in $w$ for times sufficiently close to today when dark energy is non-negligible.

\subsection{Linear perturbation results}
We now look at the physical predictions of the full linear perturbation equations derived in the previous sections. We always use the best fit parameters  from the WMAP 7-year data results \cite{Larson:2010gs}: $\Omega_{m0} = 0.265$, $n_s = 0.963$, $H_0 = 100h \rm{\ km/s/Mpc}$ ($h = 0.71$), $\Omega_k = 0$, where $n_s$ and $\Omega_k$ are the spectral index and the fractional energy density associated with the spatial curvature. These values are obtained for a $\Lambda$CDM model but may be modified once a Galileon gravity cosmology is assumed. However, for the purposes of our analysis of linear perturbations, it is sufficient to consider these values and we will provide a revised fit of the WMAP 7-year data in Galileon cosmology in a future work. The amplitude of the primordial curvature perturbations is $\Delta^2_{\mathcal{R}}(k_0)=2.43\times10^{-9}$ at a pivot scale $k_0=0.002 \rm{Mpc}^{-1}$. We set the initial conditions of the Galileon perturbation $\gamma$ and its time derivative to be zero, and have checked that the evolution of $\gamma$ is insensitive to the exact initial values.

As a consistency test of the results that follow, we checked that the perturbed quantities we obtain from {\tt CAMB} satisfy the $k$-space conservation equations, Eqs.~(\ref{conservationk1}, \ref{conservationk2}).

\subsubsection{CMB}

\begin{figure}
	\centering
	\includegraphics[scale=0.44]{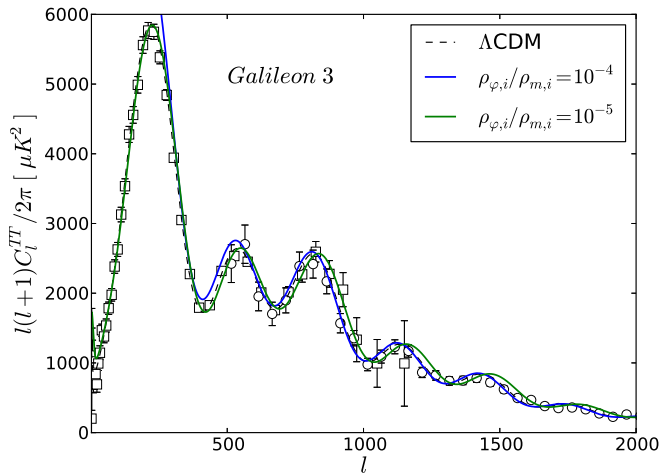}
	\caption{(Color online) CMB temperature power spectra for the Galileon 3 model with two different initial conditions and for $\Lambda$CDM (dashed black), together with the WMAP 7-year (squares) \cite{Komatsu:2010fb} and ACT (circles) \cite{Dunkley:2010ge} data. From top to bottom, at $l = 500$, the Galileon lines (solid) correspond to $\rho_{\varphi, i}/\rho_{m,i} = \{10^{-4}, 10^{-5}\}$, respectively.}
\label{rhopiorhom-Cl-gali3}\end{figure}

\begin{figure*}
	\centering
	\includegraphics[scale=0.53]{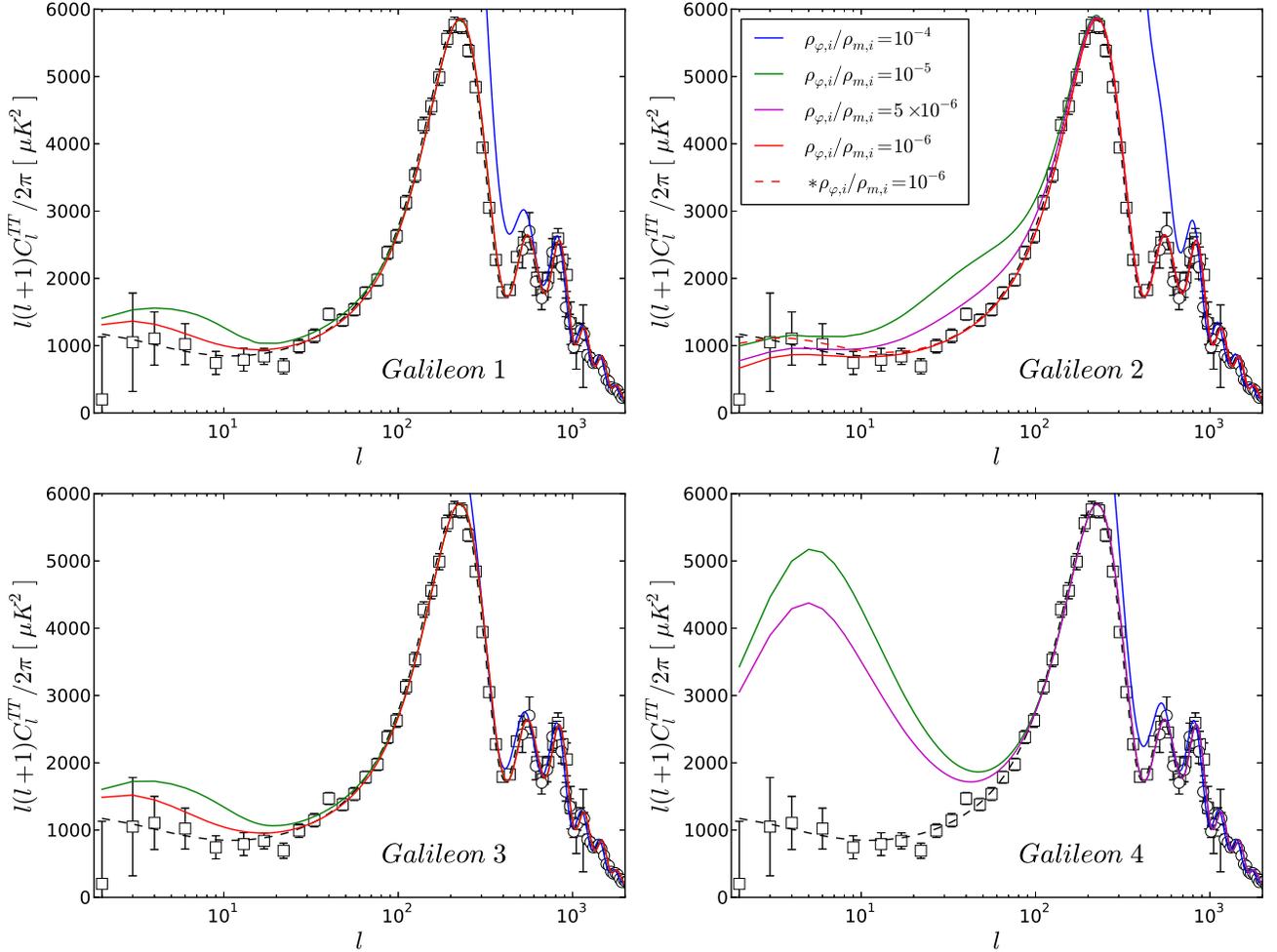}
	\caption{{(Color online) CMB power spectra for the four Galileon models for different initial conditions and $\Lambda$CDM, together with the WMAP 7-year data (squares) \cite{Komatsu:2010fb} and ACT (circles) \cite{Dunkley:2010ge} data.  In the Galileon 1 and Galileon 3 panels, from top to bottom, at $l = 10$, the lines correspond, respectively, to $\rho_{\varphi, i}/\rho_{m,i} = \{10^{-4}(\rm{not\ visible}),10^{-5},10^{-6}\}, \Lambda\rm{CDM}$. The same for the Galileon 2 and Galileon 4 panels, but for $l = 2$ and for $\rho_{\varphi, i}/\rho_{m,i} = 10^{-4}(\rm{not\ visible}), \Lambda\rm{CDM}, \rho_{\varphi, i}/\rho_{m,i} = \{*10^{-6}, 10^{-5},5\times10^{-6}, 10^{-6}\}$, and $\rho_{\varphi, i}/\rho_{m,i} = \{10^{-4}(\rm{not\ visible}),10^{-5},5\times10^{-6}\}, \Lambda\rm{CDM}$, respectively.}}
\label{rhopiorhom-Cl}\end{figure*}

\begin{figure*}\label{phi}
	\centering
	\includegraphics[scale=0.53]{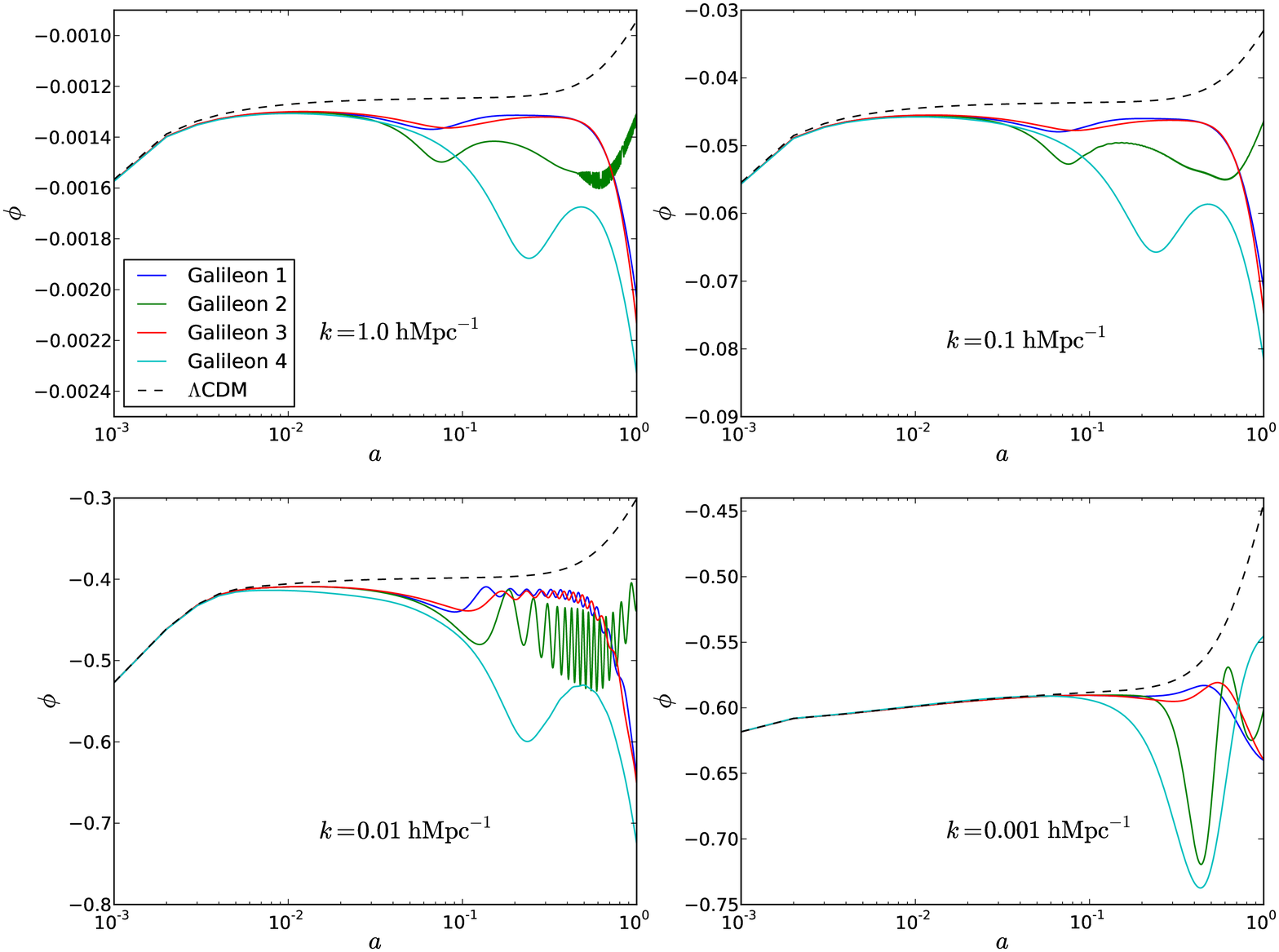}
	\caption{{(Color online) Time evolution of the Weyl gravitational potential $\phi$ for the four Galileon models and $\Lambda$CDM (dashed) for $k = \{1.0,\ 0.1,\ 0.01\  {\rm and}\  0.001\} \ \rm{hMpc}^{-1}$. All the models have the initial condition $\rho_{\varphi, i}/\rho_{m,i} = 10^{-5}$. At $a = 0.1$, for the $k = \{1.0,0.1,0.01\} \ \rm{hMpc}^{-1}$ panels, and at $a = 0.4$ for the $k = 0.001 \ \rm{hMpc}^{-1}$ panel,  from top to bottom the lines correspond, respectively, to $\Lambda$CDM, Galileon 1, Galileon 3, Galileon 2 and Galileon 4.}}
\label{phi}\end{figure*}

In Figure \ref{rhopiorhom-Cl-gali3} we plot the CMB power spectrum for the Galileon 3 model and $\Lambda$CDM  together with the WMAP 7-year \cite{Komatsu:2010fb} (squares) and ACT \cite{Dunkley:2010ge} (circles) data. Figure~\ref{rhopiorhom-Cl} is the same as Figure~\ref{rhopiorhom-Cl-gali3} but for the four models of Table \ref{Models} with a log-scaled x-axis which highlights the low-$l$ region. The effect of the Galileon field in the CMB power spectrum is mainly two-fold.  

Firstly, the modifications of the expansion rate can shift the positions of the CMB acoustic peaks. The value of the initial condition has an impact on the background expansion rate and hence on the distance to the surface of last scattering, which translates into different positions for the peaks. For sufficiently small values of $\rho_{\varphi, i}/\rho_{m,i} \lesssim 10^{-5}$ (not plotted in Figure \ref{rhopiorhom-Cl-gali3} since they are indistinguishable from the $\rho_{\varphi, i}/\rho_{m,i} = 10^{-5}$ case) the Galileon 3 curves have essentially the same peaks as a result of the almost identical expansion rate (c.f. Figure \ref{background}). The same applies for the other models Galileon 1, Galileon 2 and Galileon 4.

Secondly, the late time evolution of the gravitational potential can be also different from $\Lambda$CDM, resulting in a modified signal of the ISW effect on the largest angular scales (low $l$ in Figure~\ref{rhopiorhom-Cl}). For instance, the choice $\rho_{\varphi,i}/\rho_{m,i} = 10^{-4}$ is completely ruled out for all the models shown, since the spectrum at low $l$ is larger than the observational data by several orders of magnitude. In this case, the ISW effect is so pronounced that it dominates over the first acoustic peak and can also have an impact on the second and third ones.

Lowering the initial amount of dark energy helps to reconcile the models with the data. However, for Galileon 4 there is still too much power on large scales. Note that this model differs from Galileon 3 by having a non-vanishing value of $c_G$ and it is impossible to keep lowering the initial Galileon density ($\rho_{\varphi,i}/\rho_{m,i} \sim 5\times10^{-6}$) as theoretical instabilities start to appear. This may be a hint that the strength of the derivative coupling $c_G$ can have a crucial impact on the predictions. For all the other models (Galileon 1 to Galileon 3), for sufficiently small values of  $\rho_{\varphi,i}/\rho_{m,i}$, the dependence on the initial conditions become less pronounced and the fit to the CMB improves. There are still differences from the best fit $\Lambda$CDM model and from the data at low $l$ but since the errorbars are also larger due to cosmic variance, Galileon 1 to Galileon 3 models are still compatible with the observations. 

It is interesting to note that the CMB power spectrum for the Galileon 1 and Galileon 3 models can be quite similar although their $c_3$ and $c_4$ parameters are different. This shows that there are, to some extent, degeneracies in the Galileon model parameter space. On the other hand, changing only one of the Galileon parameters can also change considerably the CMB predictions. For instance, in the top-right panel we plot the CMB power spectrum of a model sharing all the parameters of Galileon 2 in Table \ref{Models} except that $c_4 = -1.659$, for $\rho_{\varphi,i}/\rho_{m,i} = 10^{-6}$ (dashed red). Note that $c_2$ also differs because it is tuned to yield the required amount of dark energy today, giving $c_2 = -14.968$. We see that by changing only $c_4$ the predicted CMB spectrum gets considerably closer to the data for the lowest values of $l$. It is also interesting to note that all the models have the value of $c_5$ fixed and we expect a richer phenomenology if we allow this parameter to vary as well.

To further understand the CMB predictions of the Galileon model at low $l$, we plot in Figure \ref{phi} the time evolution of the Weyl potential, $\phi$, which is the relevant quantity for the ISW effect. We show the evolution for different values of $k$ for the initial condition $\rho_{\varphi,i}/\rho_{m,i} = 10^{-5}$. The variety of evolutions can be very rich within the parameter space of the Galileon model and depends on the scale under consideration. The evolution of $\phi$ agrees, to some extent, with the $\Lambda$CDM model during the radiation dominated era. However, in the matter era, while $\phi$ is constant in the $\Lambda$CDM model, that is not the case for Galileon gravity and the gravitational potential does evolve with time. In particular, we note a very pronounced variation with time of $\phi$ for Galileon 4 during the matter era and today which explains why there is so much power at low $l$ in this model (c.f. Figure \ref{rhopiorhom-Cl}). Moreover, for the models shown, the gravitational potential suffers an overall deepening with time \cite{Appleby:2011aa, Appleby:2012ba, DeFelice:2010as, DeFelice:2011hq}, in clear contrast with the $\Lambda$CDM model where the gravitational potential gets shallower with the onset of the accelerated expansion.

\subsubsection{Weak lensing power spectrum}

The weak lensing signal of the CMB anisotropies is determined by the lensing potential $\psi$, which is an effective potential obtained by integrating the Weyl potential, $\phi$, from today to the time of last scattering \cite{Lewis:2006fu} (see also \cite{Zhao:2008bn} for a concise description and application to modified gravity theories). 

\begin{figure*}
	\centering
	\includegraphics[scale=0.53]{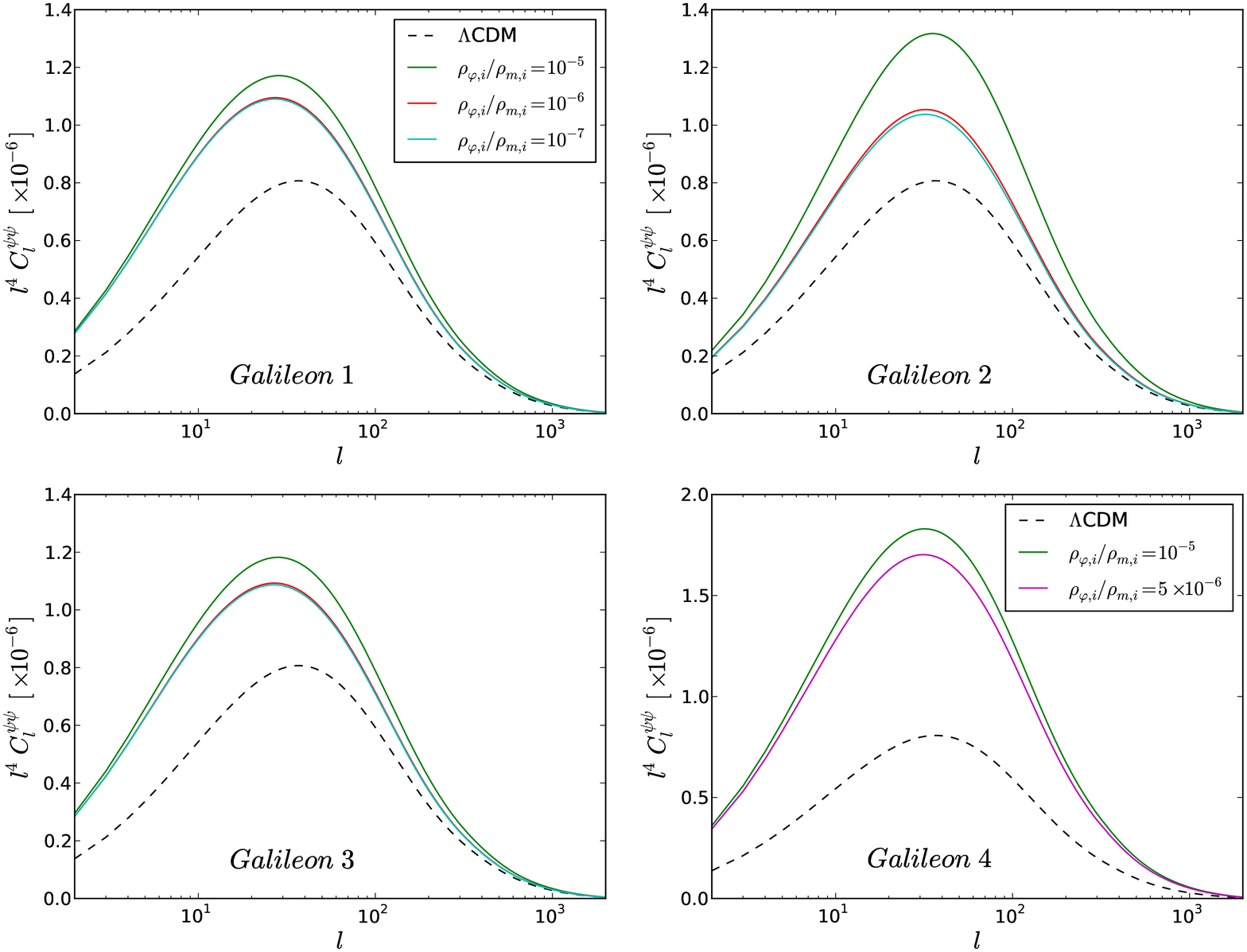}
	\caption{(Color online) Angular power spectrum of the weak lensing potential $\psi$ for the four Galileon models with different initial conditions and $\Lambda$CDM (dashed). In the Galileon 1, Galileon 2 and Galileon 3 panels, from top to bottom, the lines correspond, respectively, to $\rho_{\varphi, i}/\rho_{m,i} = \{10^{-5},10^{-6},10^{-7}\}$ and $\Lambda$CDM (the two smallest initial conditions are nearly indistinguishable in the Galileon 1 and Galileon 3 panels). The same for the Galileon 4 panel but for $\rho_{\varphi, i}/\rho_{m,i} = \{10^{-5},5\times10^{-6}\}$ and $\Lambda$CDM.}
\label{rhopiorhom-lensing}\end{figure*}

The angular power spectrum of $\psi$ is plotted in Figure  \ref{rhopiorhom-lensing} for the four Galileon models and we see that it can be noticeably larger than the $\Lambda$CDM result on all scales, as a consequence of the pronounced time variation of $\phi$ in these models (c.f. Figure \ref{phi}). The Galileon 4 model is the one where the gravitational potential deepens the most with time and it is therefore the model with the most lensing power. The initial conditions also have an impact on the result, especially for $\rho_{\varphi, i}/\rho_{m,i} \gtrsim 10^{-6}$. For instance, for the case $\rho_{\varphi, i}/\rho_{m,i} = 10^{-4}$ (which is not plotted) the power is higher by several orders of magnitude for all the models.

This is an important result and it shows that weak lensing measurements have the capability to place strong constraints on the Galileon gravity model. In particular, the Galileon 1 to Galileon 3 models, which have CMB temperature power spectrum predictions similar to that of $\Lambda$CDM for $\rho_{\varphi, i}/\rho_{m,i} = 10^{-6}$ (red line), nevertheless have very distinctive predictions for the power spectrum of the lensing potential.

\subsubsection{Matter power spectrum}

\begin{figure*}\label{rhopiorhom-Pk}
	\centering
	\includegraphics[scale=0.53]{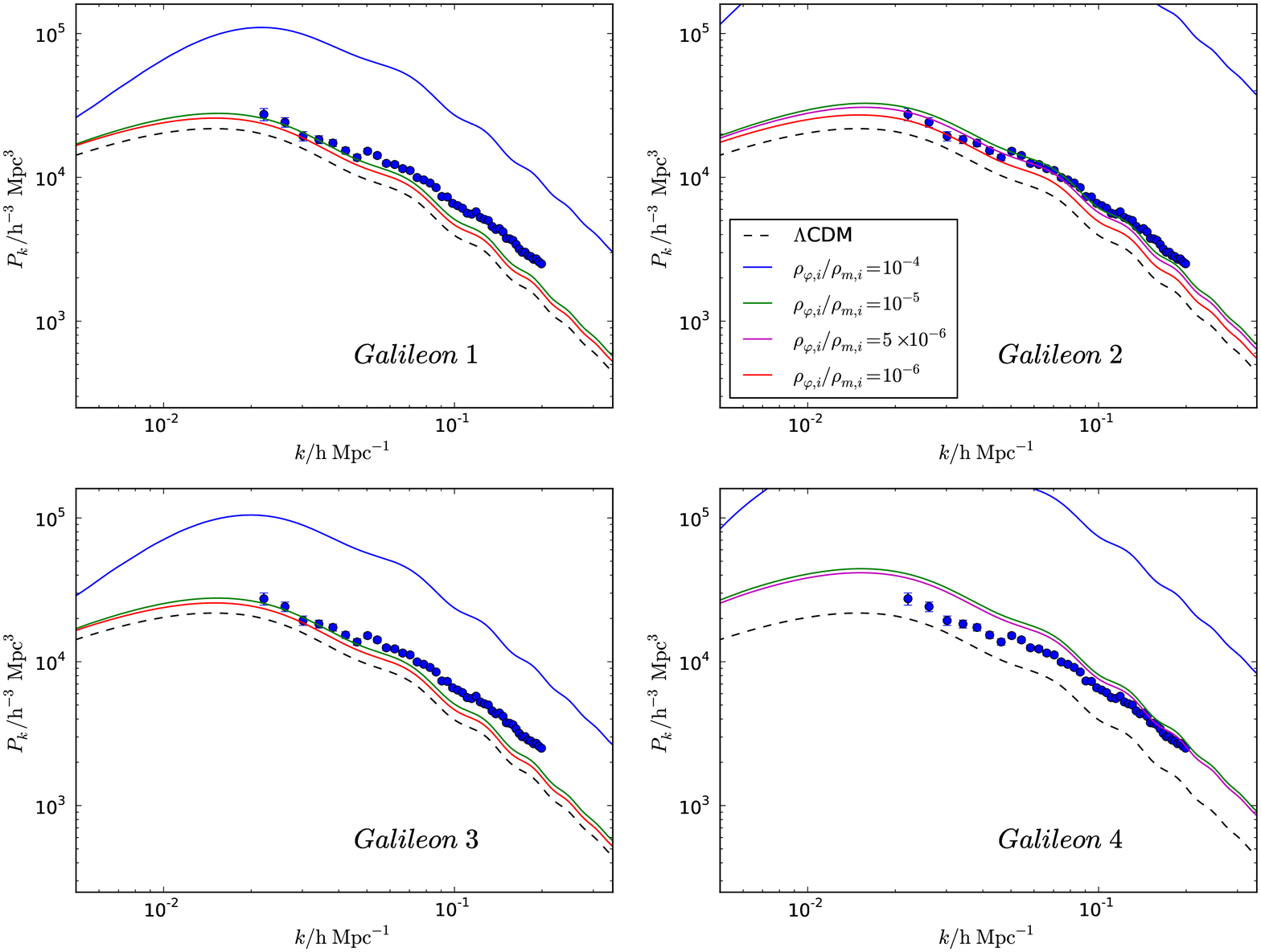}
	\caption{(Color online) Matter power spectrum at redshift $\bar{z}_{LRG} = 0.31$ for the four Galileon models with different initial conditions and $\Lambda$CDM (dashed), together with the SDSS-DR7 LRG host halo power spectrum \cite{Reid:2009xm}. $\bar{z}_{LRG}$ is the mean redshift of the LRG sample. In the Galileon 1 and Galileon 3 panels, from top to bottom, the lines correspond, respectively, to $\rho_{\varphi, i}/\rho_{m,i} = \{10^{-4},10^{-5},10^{-6}\}$ and $\Lambda\rm{CDM}$. The same for the Galileon 2 and Galileon 4 panels panels but for $\rho_{\varphi, i}/\rho_{m,i} = \{10^{-4},10^{-5},5\times10^{-6}, 10^{-6}\}, \Lambda\rm{CDM}$ and $\rho_{\varphi, i}/\rho_{m,i} = \{10^{-4},10^{-5},5\times10^{-6}\}, \Lambda\rm{CDM}$, respectively.}
\label{rhopiorhom-Pk}\end{figure*}

Figure \ref{rhopiorhom-Pk} shows the linear matter power spectrum predicted in the different models. We have chosen to plot the power spectra at redshift $\bar{z}_{LRG} = 0.31$, which is the median redshift of luminous red galaxies (LRGs) in DR7 from the Sloan Digital Sky Survey (SDDS) \cite{Eisenstein:2001}. A recent estimate of the power spectrum of LRGs is shown by the points with errorbars reproduced in each panel \cite{Reid:2009xm}. By plotting the matter power spectrum at the same redshift as the measurement, there is no need to make any adjustment for the growth factor to compare theory to observation. However, since we are plotting the prediction of linear perturbation theory in real space, there are three effects which could be responsible for any discrepancies between the theoretical spectra and the measurement: 1) Galaxy bias. This is generally modelled as a constant shift in the amplitude of the power spectrum on large scales, though simulations show that the bias is scale dependent, particularly for highly clustered objects \cite{Angulo:2008}. 2) Redshift-space distortions. Using peculiar velocities to infer the radial distance to a galaxy introduces a systematic shift in the clustering amplitude. Again, this can be scale dependent \cite{Jennings:2011}. 3) Non-linear effects. This includes the familiar mode coupling between fluctuations on different scales, but also, in the case of the Galileon models, possible screening effects which could introduce scale dependent departures from the linear perturbation theory predictions.

There are different lines of evidence which point to LRGs being biased tracers of the dark matter distribution. Interpretations of the measured clustering of LRGs in terms of empirical halo occupation distribution models suggest that these galaxies reside in massive dark matter haloes, with an effective host halo mass of $\approx 10^{14} M_{\odot}$ \cite{Wake:2008, Zheng:2009, Sawangwit:2011}. At the median redshift of the LRGs, this suggests a linear bias factor of $b \sim 2$. Measurements of the three point correlation function of LRGs can be used to infer their bias, and also return $b \sim 2$ \cite{Kulkarni:2007, Marin:2011}. For such a high bias, the amplitude boost from redshift distortions on large scales is expected to be modest. LRGs are therefore expected to have a clustering amplitude that is approximately four times higher than that of the dark matter on large scales. The measured power spectrum plotted in Figure \ref{rhopiorhom-Pk} is an estimate of the power spectrum of the {\it haloes} which host LRGs, and is not directly comparable with the estimates of the LRG bias factor outlined above. Reid et~al. \cite{Reid:2009xm} processed the LRG density field by ``collapsing'' LRGs in common dark matter haloes, to reduce the small-scale ``fingers of God'' redshift space distortion. Hence, massive haloes which host more than one LRG are given the same weight as a halo which hosts one LRG. Therefore, the effective bias of a sample of haloes weighted in this way will be smaller than the effective bias when retaining the weighting of the number of LRGs observed. If we compare the $\Lambda$CDM power spectrum to the measurement in Figure \ref{rhopiorhom-Pk}, we see that the effective bias of this sample is closer to $ b \sim \sqrt{2}$.
  
Nevertheless, despite this complication, it seems reasonable to demand that in a viable model, the observed power spectrum of LRG host haloes should have a higher amplitude than the linear theory matter power spectrum. This simple requirement puts many of the Galileon model power spectra plotted in Figure \ref{rhopiorhom-Pk} at odds with the observed power spectrum. For these models, the success of the comparison with the data depends sensitively on the value of $\rho_{\varphi,i}/\rho_{m,i}$. For instance, the initial condition $\rho_{\varphi,i}/\rho_{m,i} = 10^{-4}$ has an excess of power clearly incompatible with the observations, as it would require a bias parameter $b \ll 1$. Lowering $\rho_{\varphi,i}/\rho_{m,i}$ allows a better match to the observations to be obtained and the results become less sensitive to the initial conditions (lower initial conditions have nearly the same prediction as $\rho_{\varphi,i}/\rho_{m,i} = 10^{-6}$). However, all the models still produce an excess of power when compared to $\Lambda$CDM indicating that the formation of linear structure can be highly enhanced by the modifications of gravity in the Galileon model, a conclusion in agreement with previous linear perturbation studies in the literature \cite{DeFelice:2010as, Appleby:2011aa, Appleby:2012ba, Okada:2012mn}.  The Galileon 4 model is the one with the worst fit, even for the lowest initial condition $\rho_{\varphi,i}/\rho_{m,i} = 5\times10^{-6}$ (recall that in this model lower initial conditions lead to the appearance of instabilities). This indicates once again that the $c_G$ parameter can have a critical impact on the results. For $k \lesssim 0.05 \ \rm{hMpc}^{-1}$, all the other models would agree very well with the data if $b = 1$ and $\rho_{\varphi,i}/\rho_{m,i} \lesssim 10^{-5}$. However, considering $b > 1$ will increase the power on all scales, which could in principle be used to place strong observational contraints on Galileon models.

We should stress, however, that when comparing the Galileon model with $\Lambda$CDM and clustering data one should be cautious about the validity of linear perturbation theory since the scale at which the Vainshtein screening effect becomes important is not well known. For example, numerical simulations have shown that in other modified gravity models such as the $f(R)$ and dilaton \cite{Brax:2010gi, Brax:2011ja, Brax:2012nk}, linear perturbation theory can fail even on scales as large as $k \sim 0.01 \ \rm{hMpc}^{-1}$ because of the screening \cite{Brax:2012nk, Li:2012by}. {As a result, a detailed study of the effects of the Vainshtein screening is necessary for a more complete comparison of the theory predictions against the observations. This is beyond the scope of the present paper and will be left for future work (see however \cite{Bellini:2012qn, Kimura:2011dc, Babichev:2011iz, DeFelice:2011th, Hiramatsu:2012xj, Burrage:2010rs, Iorio:2012pv} for work already taken in this direction).}

\subsubsection{Clustering of the Galileon field}

\begin{figure*}
	\centering
	\includegraphics[scale=0.53]{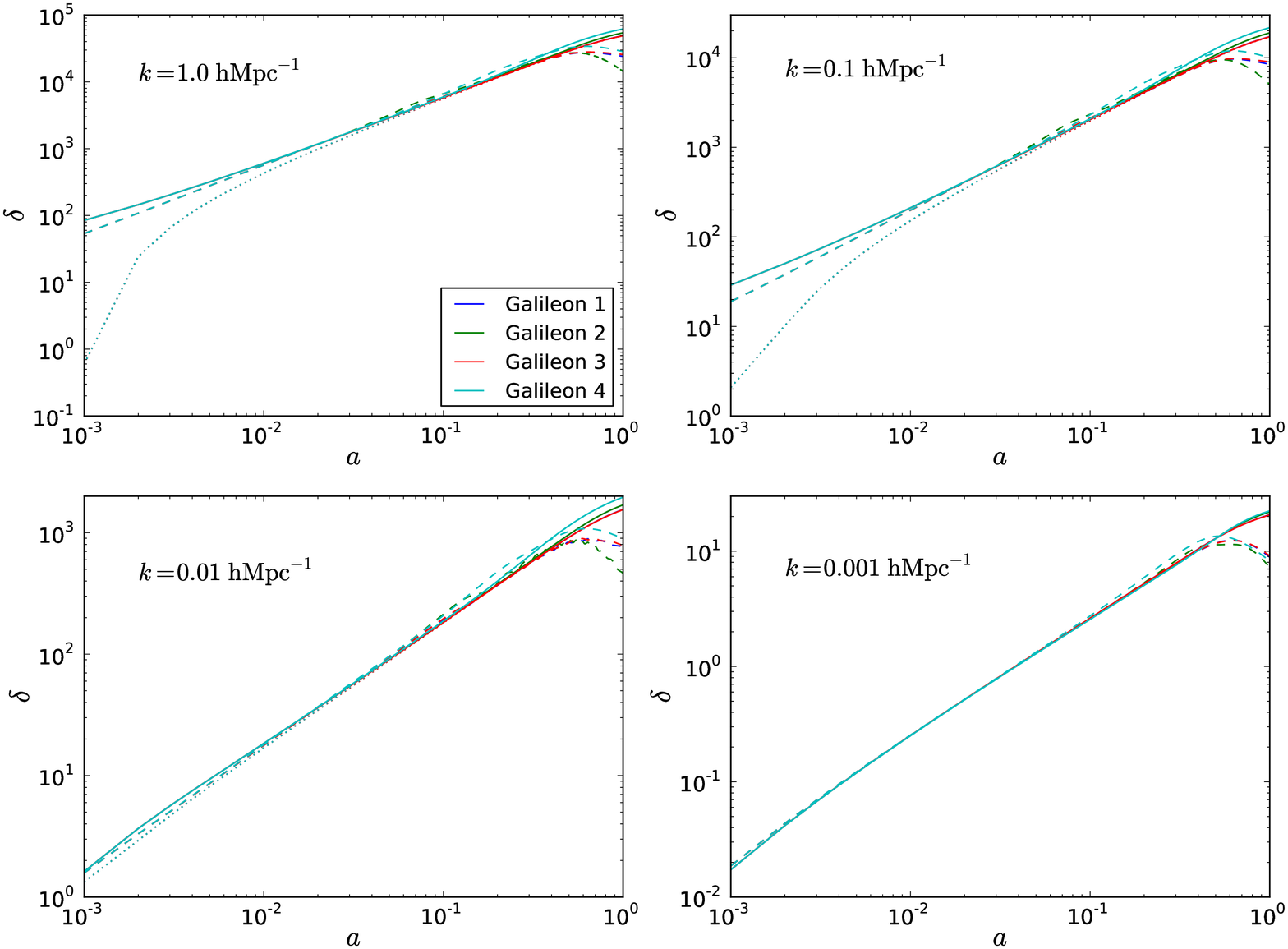}
	\caption{(Color online) {Time evolution of the linear density contrast of dark matter (DM, solid lines), $\delta_{\rm{DM}} = {\rho_{DM}}/\bar{\rho}_{DM} - 1$, baryonic matter (B, dotted lines), $\delta_{\rm{B}} = {\rho_{B}}/\bar{\rho}_{B} - 1$, and Galileon field (dashed lines), $\delta_{\varphi} = {\rho_{\varphi}}/\bar{\rho}_{\varphi} - 1$, for the four Galileon models for $k = \{1.0,\ 0.1,\ 0.01\  \rm{and}\  0.001\} \ \rm{hMpc}^{-1}$. All the models have initial condition $\rho_{\varphi, i}/\rho_{m,i} = 10^{-5}$. All the models behave more or less in the same way for all the scales. In the $k = 0.001 \ \rm{hMpc}^{-1}$ panel, the DM and B lines are indistinguishable.}}
\label{density-contrast}\end{figure*}

We now turn the attention to the time evolution of the linear density contrast {of the Galileon field} $\delta_{\varphi} = {\rho_{\varphi}}/\bar{\rho}_{\varphi} - 1$. This is plotted in Figure \ref{density-contrast} for the initial condition $\rho_{\varphi, i}/\rho_{m,i} = 10^{-5}$. We see that the Galileon density contrast (dashed lines) can be large, being comparable with the dark matter (solid lines) and baryonic matter (dotted lines) density contrasts throughout most of the evolution. This happens for all the scales considered including small scales such as $k = 1.0 \ \rm{hMpc}^{-1}$.

This strong clustering of the Galileon field has a large impact on the evolution of the Weyl gravitational potential $\phi$ which directly determines many observables such as the ISW effect (c.f.~Figure \ref{rhopiorhom-Cl}), weak lensing (c.f.~Figure \ref{rhopiorhom-lensing}) and clustering of matter (c.f.~Figure \ref{rhopiorhom-Pk}).

One can also note that the Galileon density contrast starts to decrease with time close to the present day. This could be due to the rapid growth of the Galileon background density at those times ($w < -1$) which leads to a decrease of $\delta_\varphi = \rho_\varphi / \bar{\rho}_\varphi - 1$.

\subsubsection{Quasi-static limit approximation}

\begin{figure*}
	\centering
	\includegraphics[scale=0.53]{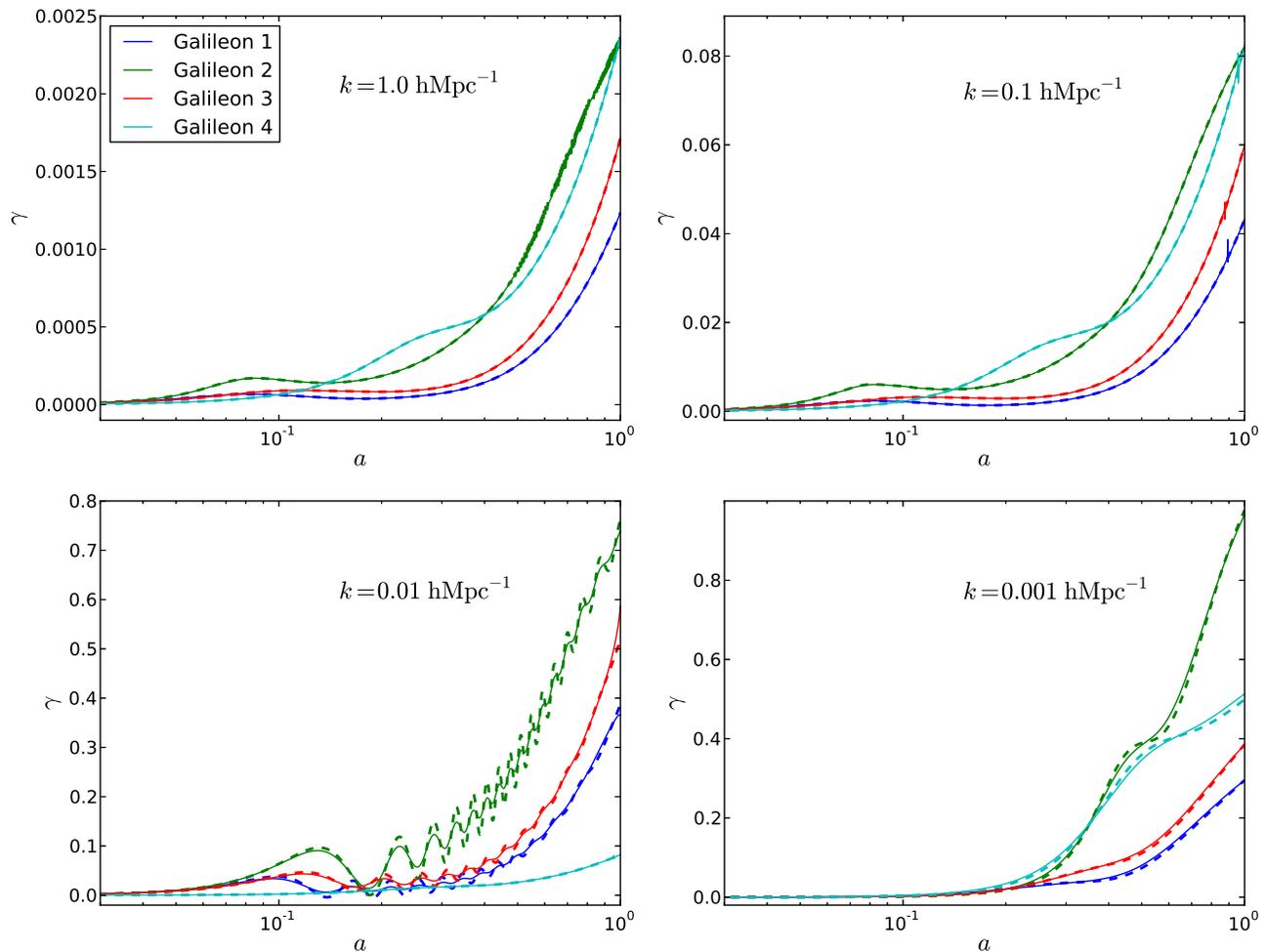}
	\caption{{(Color online) Time evolution of the $k$-space Galileon field perturbation $\gamma$ (dashed) along with the corresponding quasi-static limit (solid), for the four Galileon models for $k = \{1.0,\ 0.1,\ 0.01\  \rm{and}\  0.001\} \ \rm{hMpc}^{-1}$. All the models have initial condition $\rho_{\varphi, i}/\rho_{m,i} = 10^{-5}$. For the $k = 1.0 \ \rm{hMpc}^{-1}$ and $k = 0.1 \ \rm{hMpc}^{-1}$ panels at $a = 0.2$, from top to bottom, the lines correspond to the models Galileon 4, Galileon 2, Galileon 3 and Galileon 1, respectively. For the $k = 0.01 \ \rm{hMpc}^{-1}$ panel at $a = 1.0$, from top to bottom the lines correspond to Galileon 2, Galileon 3, Galileon 1 and Galileon 4, respectively. For the $k = 0.001 \ \rm{hMpc}^{-1}$ panel at $a = 1.0$, from top to bottom, the lines correspond to Galileon 2, Galileon 4, Galileon 3 and Galileon 1, respectively.}}
\label{gamma}\end{figure*}

In Figure~\ref{gamma} we plot the time evolution of the $k$-space Galileon perturbation, $\gamma$ (dashed), along with the corresponding solution obtained in the quasi-static limit (solid). The quasi-static limit is the limit in which the spatial derivatives of the field are dominant over the time derivative ones. Practically, this means neglecting all terms in the field equations that are suppressed by $\mathcal{H}^2/k^2$ or $\varphi'/k^2$.

As for the evolution of the density contrast $\delta_\varphi$ and the Weyl potential $\phi$, here there is also a strong scale dependence. Moreover, we see that even for near-horizon scales such as $k = 0.001 \ \rm{hMpc}^{-1}$ the quasi-static limit can be a good (though not perfect) approximation to the full solution. In particular, in the Galileon 2 curves with $k = 0.01 \ \rm{hMpc}^{-1}$, one can see that the quasi-static approximation agrees quite well with the full solution despite the oscillations in the latter.

The quasi-static limit appears therefore to be valid for many cases in the Galileon model, especially when one is interested in subhorizon scales. However, it breaks down on superhorizon scales and can lead to inaccurate predictions of the ISW effect and weak lensing signals. Moreover, as we can see from the lower-right panel of Figure~\ref{gamma}, on near-horizon scales with $k = 0.001 \ \rm{hMpc}^{-1}$, the error of this approximation can be a few percent, which is much larger than the numerical error of the {\tt CAMB} code (which is at sub-percent level). For these reasons, we prefer to use the full numerical solution in the modified {\tt CAMB} code.

\section{Conclusion}\label{Conclusion}

We have studied the cosmology of Galileon gravity models at the linear perturbation level. For this we derived the full CGI perturbation equations using two independent methods: the normal procedure of linearising the full field equations and an alternative derivation that is particularly suitable for models like Galileon gravity, where the shape of the Lagrangian is fixed by certain symmetries (e.g., there are no free functions such as the potential in quintessence and $f(R)$ gravity models) and the field equations only contain up to second-order derivatives. The second derivation is particularly appealing because it is much simpler than the first one, which is very lengthy and complicated for the full Galileon model. We checked that the two methods give the same set of perturbation equations, and then solved these equations using a modified version of the {\tt CAMB} code, which we tested by performing several successful consistency tests.

Our code also solves the background expansion history in Galileon models and our results agree with those in the literature. We find that the expansion rate in Galileon cosmology can depend sensitively on the initial value of the Galilean energy density, especially if the latter is not small, e.g., if $\rho_{\varphi, i} / \rho_{m,i} \gtrsim 10^{-5}$. Throughout the evolution, the expansion rate can be faster or slower than in $\Lambda$CDM and the Galileon equation-of-state parameter can cross the phantom line ($w<-1$) in a way which is free of ghost-like instabilities. 

The modified background expansion translates into a different age of the Universe and distance to last scattering, which leads to a visible shift in the positions of the acoustic peaks of the CMB temperature power spectrum.
The strongest effect of the Galileon field on the CMB temperature power spectrum, however, appears to be on the largest angular scales (low values of $l$), where the full power receives a significant contribution from the integrated Sachs-Wolfe effect, which is due to the late-time evolution of the gravitational potential $\phi$. Indeed, we found that in Galileon models the gravitational potential evolves even during the matter dominated era and can undergo an overall deepening at late times. This is very different from the standard $\Lambda$CDM prediction that the gravitational potential is constant during matter domination and becomes shallower when the expansion of the Universe starts to accelerate. The origin of the abnormal evolution in the gravitational potential can be traced back to the pressure perturbation and anisotropic stress of the Galileon field, which cause it to cluster strongly (comparable to the clustering of dark and baryonic matter species) on all scales.  

The evolution of the gravitational potential influences a number of cosmological observables, both directly and indirectly. In addition to the ISW effect, it also has strong impact on the growth of matter density perturbations (and therefore the linear and nonlinear matter power spectra), weak gravitational lensing and their cross correlations. In particular, we have shown that the Galileon model can predict considerably more power than $\Lambda$CDM for the weak lensing power spectrum at all scales, even if their predictions for the CMB power spectrum more or less agree. Galileon models might also have distinctive predictions for the cross correlation of the ISW effect with the galaxy distribution. These are important observational signatures in the linear perturbation regime that can in principle help to distinguish the Galileon models from the standard $\Lambda$CDM paradigm.

On the other hand, the sensitive dependence of the Galileon behaviour on the model parameters makes the phenomenology of the Galileon cosmology especially rich. For example, by tuning the parameters in the Galileon Lagrangian, one can get a CMB power spectrum which is very close to the $\Lambda$CDM prediction and therefore hard to distinguish by looking at very large scales.

On subhorizon scales, we have seen that the linear growth of matter density perturbations  can be significantly enhanced with respect to the $\Lambda$CDM results, even for those model parameters that lead to similar CMB power spectrum. However, in Galileon models, the Vainshtein screening mechanism is at play and its potential influence on the clustering of matter is still to be properly understood. As an analogy, in other modified gravity models such as the $f(R)$ gravity, the chameleon screening effect has been shown to make the linear perturbation theory a poor approximation even on scales as large as $k = 0.01 \ \rm{hMpc}^{-1}$. We therefore conclude that a better understanding of the true impact of the Vainshtein screening is necessary, before attempting a more rigorous confrontation of the predicted matter power spectrum with measurements of galaxy clustering. Such a study will be left for future work.

Finally, we have seen that the quasi-static approximation for the evolution of the Galileon field perturbation serves as a good approximation on subhorizon scales for the models we have shown in this paper. It works reasonably well on near-horizon scales such as $k = 0.001 \ \rm{hMpc}^{-1}$, with an error of the order of a few percent. However, for accuracy considerations we solve the full evolution equation of the Galileon perturbation in our code, which does not take much longer anyway.

In conclusion, we have shown that the detailed study of the full perturbation equations unveils a rich phenomenology in Galileon gravity models. The full cosmological parameter space increases considerably in Galileon gravity. Even with current data, the indications are that strong constraints can be placed on this parameter space. In a future project we will use our modification to the {\tt CAMB} software to carry out a formal study of the Galileon parameter space.

\begin{acknowledgments}

We are grateful to Gong-Bo Zhao, {Alexander Vikman, Antony Lewis, Pedro Ferreira and Mustafa Amin} for useful comments. AB acknowledges support by FCT-Portugal through grant SFRH/BD/75791/2011. BL is supported by the Royal Astronomical Society and Durham University. SP thanks the Galileo Galilei Institute for Theoretical Physics for its hospitality during part of this study.

\end{acknowledgments}

\clearpage

\appendix

\section{The Covariant Field Equations in Galileon Gravity}

\label{Appendix A}
The modified Einstein field equations and the Galileon field equation of motion can be obtained by varying the action with respect to $g_{\mu\nu}$ and $\varphi$, respectively. The Einstein field equations are given by:
\bq
G_{\mu\nu} = \kappa && \left[T_{\mu\nu}^f + T_{\mu\nu}^{c_2} + T_{\mu\nu}^{c_3}  + T_{\mu\nu}^{c_4} + T_{\mu\nu}^{c_5} + T_{\mu\nu}^{c_G}\right],\ \ 
\eq
where 
\begin{widetext}
\bq
T_{\mu\nu}^{c_2} &=& c_2\left[ \nabla_\mu\varphi\nabla_\nu\varphi - \frac{1}{2}g_{\mu\nu}\nabla^\alpha\varphi\nabla_\alpha\varphi \right], \\ 
T_{\mu\nu}^{c_3} &=& \frac{c_3}{M^3}\left[ 2\nabla_\mu\varphi\nabla_\nu\varphi\Box\varphi + 2g_{\mu\nu}\nabla_\alpha\varphi\nabla_\beta\varphi\nabla^\alpha\nabla^\beta\varphi - 4\nabla^\lambda\varphi\nabla_{(\mu}\varphi\nabla_{\nu)}\nabla_\lambda\varphi \right], \\
T_{\mu\nu}^{c_4} &=& \frac{c_4}{M^6}g_{\mu\nu} \left[ (\Box\varphi)^2\nabla_\lambda\varphi\nabla^\lambda\varphi - \frac{1}{12}R\left(\nabla_\alpha\varphi\nabla^\alpha\varphi\right)^2 + 4\Box\varphi\nabla^\alpha\varphi\nabla^\beta\varphi\nabla_\alpha\nabla_\beta\varphi - 4\nabla_\lambda\nabla_\alpha\varphi\nabla^\lambda\nabla_\beta\varphi\nabla^\alpha\varphi\nabla^\beta\varphi \right. \nonumber \\ 
&& \ \ \ \ \ \ \ \ \ \ \ \ \ \ \ \ \ \left. - \nabla_\lambda\varphi\nabla^\lambda\varphi\nabla_\alpha\nabla_\beta\varphi\nabla^\alpha\nabla^\beta\varphi - R_{\alpha\beta}\nabla^\alpha\varphi\nabla^\beta\varphi\nabla_\lambda\varphi\nabla^\lambda\varphi\right] \nonumber \\
&& + \frac{c_4}{M^6}\left[ 2(\Box\varphi)^2\nabla_\mu\varphi\nabla_\nu\varphi + 2\nabla_\lambda\varphi\nabla^\lambda\varphi\nabla^\rho\varphi R_{\rho(\mu}\nabla_{\nu)}\varphi - 8\Box\varphi\nabla^\lambda\varphi\nabla_\lambda\nabla_{(\mu}\varphi\nabla_{\nu)}\varphi \right. \nonumber \\
&& \ \ \ \ \ \ \ \ \ \ \ \ \left. -2\nabla_\alpha\nabla_\beta\varphi\nabla^\alpha\nabla^\beta\varphi\nabla_\mu\varphi\nabla_\nu\varphi + 8\nabla^\lambda\varphi\nabla_\rho\nabla_\lambda\varphi\nabla^\rho\nabla_{(\mu}\varphi\nabla_{\nu)}\varphi - 2\Box\varphi\nabla_\lambda\varphi\nabla^\lambda\varphi\nabla_\mu\nabla_\nu\varphi \right. \nonumber \\
&& \ \ \ \ \ \ \ \ \ \ \ \ \left. -4\nabla_\alpha\nabla_\beta\varphi\nabla^\alpha\varphi\nabla^\beta\varphi\nabla_\mu\nabla_\nu\varphi - \frac{2}{3}R\nabla_\lambda\varphi\nabla^\lambda\varphi\nabla_\mu\varphi\nabla_\nu\varphi + \frac{1}{2}R_{\mu\nu}\left(\nabla_\alpha\varphi\nabla^\alpha\varphi\right)^2 \right. \nonumber \\
&& \ \ \ \ \ \ \ \ \ \ \ \ \left. + 2\nabla_\mu\nabla_\alpha\varphi\nabla_\nu\nabla^\alpha\varphi\nabla_\lambda\varphi\nabla^\lambda\varphi + 4\nabla_\mu\nabla_\alpha\varphi\nabla_\nu\nabla_\beta\varphi\nabla^\alpha\varphi\nabla^\beta\varphi + 2\mathcal{W}_{\mu\alpha\nu\beta}\nabla^\alpha\varphi\nabla^\beta\varphi\nabla_\lambda\varphi\nabla^\lambda\varphi\right],\\
T_{\mu\nu}^{c_5} &=&  \frac{c_5}{M^9}g_{\mu\nu} \left[ (\Box\varphi)^3\nabla_\lambda\varphi\nabla^\lambda\varphi + 3(\Box\varphi)^2\nabla^\alpha\nabla^\beta\varphi\nabla_\alpha\varphi\nabla_\beta\varphi - 3\Box\varphi\nabla_\lambda\varphi\nabla^\lambda\varphi\nabla_\alpha\nabla_\beta\varphi\nabla^\alpha\nabla^\beta\varphi \right. \nonumber \\
&& \ \ \ \ \ \ \ \ \ \ \ \ \ \ \ \left. -6\Box\varphi\nabla^\alpha\nabla^\beta\varphi\nabla_\alpha\nabla_\lambda\varphi\nabla_\beta\varphi\nabla^\lambda\varphi  + 2\nabla_\lambda\varphi\nabla^\lambda\varphi\nabla_\alpha\nabla^\beta\varphi\nabla_\beta\nabla^\gamma\varphi\nabla_\gamma\nabla^\alpha\varphi \right. \nonumber \\
&& \ \ \ \ \ \ \ \ \ \ \ \ \ \ \ \left. -3\nabla_\alpha\nabla_\beta\varphi\nabla^\alpha\nabla^\beta\varphi\nabla_\rho\nabla_\sigma\varphi\nabla^\rho\varphi\nabla^\sigma\varphi + 6\nabla_\alpha\nabla_\beta\varphi\nabla^\beta\nabla^\gamma\varphi\nabla_\gamma\nabla_\lambda\varphi\nabla^\alpha\varphi\nabla^\lambda\varphi - R\Box\varphi\left(\nabla_\lambda\varphi\nabla^\lambda\varphi\right)^2 \right. \nonumber \\ 
&& \ \ \ \ \ \ \ \ \ \ \ \ \ \ \ \left. +\frac{3}{2}R_{\alpha\beta}\nabla^\alpha\nabla^\beta\varphi\left(\nabla_\lambda\varphi\nabla^\lambda\varphi\right)^2 +3\nabla_\lambda\varphi\nabla^\lambda\varphi\nabla^\alpha\varphi\nabla^\beta\varphi\nabla^\gamma\nabla^\sigma\varphi\mathcal{W}_{\alpha\gamma\beta\sigma}\right] \nonumber \\
&& +\frac{c_5}{M^9}\left[ (\Box\varphi)^3\nabla_\mu\varphi\nabla_\nu\varphi - 3(\Box\varphi)^2\nabla_\lambda\varphi\nabla^\lambda\varphi\nabla_\mu\nabla_\nu\varphi - 6(\Box\varphi)^2\nabla^\lambda\varphi\nabla_\lambda\nabla_{(\mu}\varphi\nabla_{\nu)}\varphi + 6\Box\varphi\nabla_\lambda\varphi\nabla^\lambda\varphi\nabla_\mu\nabla_\alpha\varphi\nabla_\nu\nabla^\alpha\varphi \right. \nonumber \\
&& \ \ \ \ \ \ \ \ \ \ \ \ \ \left. -6\Box\varphi\nabla_\alpha\nabla_\beta\varphi\nabla^\alpha\varphi\nabla^\beta\varphi\nabla_\mu\nabla_\nu\varphi - 3\Box\varphi\nabla_\alpha\nabla_\beta\varphi\nabla^\alpha\nabla^\beta\varphi\nabla_\mu\varphi\nabla_\nu\varphi + 6\Box\varphi\nabla^\alpha\varphi\nabla^\beta\varphi\nabla_\mu\nabla_\alpha\varphi\nabla_\nu\nabla_\beta\varphi \right. \nonumber \\
&& \ \ \ \ \ \ \ \ \ \ \ \ \ \left. +12\Box\varphi\nabla_\alpha\varphi\nabla^\alpha\nabla^\beta\varphi\nabla_\beta\nabla_{(\mu}\varphi\nabla_{\nu)}\varphi + 3\nabla_\lambda\varphi\nabla^\lambda\varphi\nabla_\alpha\nabla_\beta\varphi\nabla^\alpha\nabla^\beta\varphi\nabla_\mu\nabla_\nu\varphi\ - 6\nabla_\lambda\varphi\nabla^\lambda\varphi\nabla^\alpha\nabla^\beta\varphi\nabla_\mu\nabla_\alpha\varphi\nabla_\nu\nabla_\beta\varphi \right. \nonumber \\
&& \ \ \ \ \ \ \ \ \ \ \ \ \ \left. +6\nabla^\alpha\nabla^\beta\varphi\nabla_\alpha\varphi\nabla_\beta\varphi\nabla_\mu\nabla_\lambda\varphi\nabla_\nu\nabla^\lambda\varphi + 6\nabla^\alpha\nabla^\lambda\varphi\nabla_\beta\nabla_\lambda\varphi\nabla_\alpha\varphi\nabla^\beta\varphi\nabla_\mu\nabla_\nu\varphi + 2\nabla_\alpha\nabla^\beta\varphi\nabla_\beta\nabla^\lambda\varphi\nabla_\lambda\nabla^\alpha\varphi\nabla_\mu\varphi\nabla_\nu\varphi \right. \nonumber \\
&&\ \ \ \ \ \ \ \ \ \ \ \ \ \left. + 6\nabla^\alpha\nabla^\beta\varphi\nabla_\alpha\nabla_\beta\varphi\nabla^\lambda\varphi\nabla_\lambda\nabla_{(\mu}\varphi\nabla_{\nu)}\varphi - 12\nabla^\alpha\varphi\nabla_\alpha\nabla_\beta\varphi\nabla^\beta\nabla^\lambda\varphi\nabla_\lambda\nabla_{(\mu}\varphi\nabla_{\nu)}\varphi \right. \nonumber \\
&&\ \ \ \ \ \ \ \ \ \ \ \ \ \left. -12\nabla_\alpha\varphi\nabla^\alpha\nabla^\lambda\varphi\nabla^\sigma\varphi\nabla_\lambda\nabla_{(\mu}\varphi\nabla_{\nu)}\nabla_\sigma\varphi + \frac{3}{2}\Box\varphi\left(\nabla_\lambda\varphi\nabla^\lambda\varphi\right)^2R_{\mu\nu} - 3\left(\nabla_\lambda\varphi\nabla^\lambda\varphi\right)^2R_{\sigma(\mu}\nabla_{\nu)}\nabla^\sigma\varphi \right. \nonumber \\
&& \ \ \ \ \ \ \ \ \ \ \ \ \ \left. + R\left(\nabla_\lambda\varphi\nabla^\lambda\varphi\right)^2\nabla_\mu\nabla_\nu\varphi + 3\Box\varphi\nabla_\lambda\varphi\nabla^\lambda\varphi\nabla^\alpha\varphi\nabla^\beta\varphi\mathcal{W}_{\mu\alpha\nu\beta}  - 6\nabla_\lambda\varphi\nabla^\lambda\varphi\nabla^\alpha\varphi\nabla^\beta\nabla^\gamma\varphi\nabla_{(\mu}\varphi\mathcal{W}_{\nu)\beta\alpha\gamma} \right. \nonumber \\
&& \ \ \ \ \ \ \ \ \ \ \ \ \ \left. +6\nabla_\lambda\varphi\nabla^\lambda\varphi\nabla^\alpha\varphi\nabla^\beta\varphi\nabla^\gamma\nabla_{(\mu}\varphi\mathcal{W}_{\nu)\alpha\beta\gamma} - 6\nabla_\lambda\varphi\nabla^\lambda\varphi\nabla_\alpha\varphi\nabla^\alpha\nabla_\beta\varphi\nabla_\gamma\varphi\mathcal{W}_{(\mu \ \ \nu)}^{\ \ \beta \ \ \gamma}\right], \\
T_{\mu\nu}^{c_G} &=& \frac{M_{\rm Pl}}{M^3}c_G \left[ g_{\mu\nu} \left( (\Box\varphi)^2 - \nabla_\alpha\nabla_\beta\varphi\nabla^\alpha\nabla^\beta\varphi\right)  + 2\nabla_\mu\nabla_\lambda\varphi\nabla_\nu\nabla^\lambda\varphi - 2\Box\varphi\nabla_\mu\nabla_\nu\varphi \right. \nonumber \\ 
&& \left. + 2R_{\lambda(\mu}\nabla_{\nu)}\varphi\nabla^\lambda\varphi  - R^{\alpha\beta}\nabla_\alpha\varphi\nabla_\beta\varphi g_{\mu\nu} + 2\mathcal{W}_{\alpha\mu\beta\nu}\nabla^\alpha\varphi\nabla^\beta\varphi - \frac{2}{3}R\nabla_\mu\varphi\nabla_\nu\varphi + \frac{1}{6}R\nabla^\lambda\varphi\nabla_\lambda\varphi g_{\mu\nu}\right].
\eq
\end{widetext}

The Galileon field equation of motion is given by:

\begin{widetext}
\bq\label{Full-EOM}
0 &=& c_2\Box\varphi + 2\frac{c_3}{M^3}\left[ (\Box\varphi)^2 - \nabla^\alpha\nabla^\beta\varphi\nabla_\alpha\nabla_\beta\varphi - R_{\alpha\beta}\nabla^\alpha\varphi\nabla^\beta\varphi\right] \nonumber \\
&& + \frac{c_4}{M^6}\left[ 2(\Box\varphi)^3 - 6\Box\varphi\nabla_\alpha\nabla_\beta\varphi\nabla^\alpha\nabla^\beta\varphi + 4\nabla_\alpha\nabla_\beta\varphi\nabla^\beta\nabla^\gamma\varphi\nabla_\gamma\nabla^\alpha\varphi - \frac{4}{3}R\nabla^\alpha\varphi\nabla^\beta\varphi\nabla_\alpha\nabla_\beta\varphi \right. \nonumber \\
&& \ \ \ \ \ \ \ \ \left. -\frac{5}{3}R\Box\varphi\nabla_\lambda\varphi\nabla^\lambda\varphi + 4R_{\alpha\beta}\nabla^\alpha\nabla^\beta\varphi\nabla_\lambda\varphi\nabla^\lambda\varphi + 4R_{\alpha\beta}\nabla^\alpha\varphi\nabla^\lambda\varphi\nabla^\beta\nabla_\lambda\varphi - 2\Box\varphi R_{\alpha\beta}\nabla^\alpha\varphi\nabla^\beta\varphi \right. \nonumber \\
&& \ \ \ \ \ \ \ \ \left. +4\mathcal{W}_{\alpha\beta\lambda\rho}\nabla^\alpha\nabla^\lambda\varphi\nabla^\beta\varphi\nabla^\rho\varphi\right] \nonumber \\
&& + \frac{c_5}{M^9}\left[(\Box\varphi)^4 - 6(\Box\varphi)^2\nabla_\alpha\nabla_\beta\varphi\nabla^\alpha\nabla^\beta\varphi + 3\left(\nabla_\alpha\nabla_\beta\varphi\nabla^\alpha\nabla^\beta\varphi\right)^2 - 6\nabla_\alpha\nabla^\rho\varphi\nabla_\beta\nabla_\rho\varphi\nabla^\alpha\nabla^\lambda\varphi\nabla^\beta\nabla_\lambda\varphi \right. \nonumber \\
&& \ \ \ \ \ \ \ \ \left. +8\Box\varphi\nabla_\alpha\nabla^\beta\varphi\nabla_\beta\nabla^\lambda\varphi\nabla_\lambda\nabla^\alpha\varphi +6\Box\varphi\nabla_\lambda\varphi\nabla^\lambda\varphi R_{\alpha\beta}\nabla^\alpha\nabla^\beta\varphi - 2R(\Box\varphi)^2\nabla_\lambda\varphi\nabla^\lambda\varphi \right. \nonumber \\
&& \ \ \ \ \ \ \ \ \left. + \frac{1}{2}RR_{\alpha\beta}\nabla^\alpha\varphi\nabla^\beta\varphi\nabla^\lambda\varphi\nabla_\lambda\varphi - 6R_{\alpha\beta}\nabla^\lambda\varphi\nabla_\lambda\varphi\nabla^\alpha\nabla^\sigma\varphi\nabla^\beta\nabla_\sigma\varphi + 2R\nabla^\lambda\varphi\nabla_\lambda\varphi\nabla^\alpha\nabla^\beta\varphi\nabla_\alpha\nabla_\beta\varphi \right. \nonumber \\
&& \ \ \ \ \ \ \ \ \left. {-\frac{3}{2}}R_{\rho\sigma}R^{\rho\sigma}\left(\nabla_\lambda\varphi\nabla^\lambda\varphi\right)^2 + {\frac{1}{4}}R^2\left(\nabla_\lambda\varphi\nabla^\lambda\varphi\right)^2 + 6\Box\varphi\mathcal{W}_{\rho\alpha\sigma\beta}\nabla^\rho\nabla^\sigma\varphi\nabla^\alpha\varphi\nabla^\beta\varphi \right. \nonumber \\
&& \ \ \ \ \ \ \ \ \left. + 12\mathcal{W}_{\rho\alpha\beta\sigma}\nabla^\rho\nabla^\sigma\varphi\nabla^\beta\nabla^\lambda\varphi\nabla^\alpha\varphi\nabla_\lambda\varphi + 3\mathcal{W}_{\rho\alpha\beta\sigma}\nabla^\rho\nabla^\sigma\varphi\nabla^\alpha\nabla^\beta\varphi\nabla^\lambda\varphi\nabla_\lambda\varphi \right. \nonumber \\
&& \ \ \ \ \ \ \ \ \left. + 6\mathcal{W}_{\alpha\rho\sigma\beta}\nabla^\rho\varphi\nabla^\sigma\varphi\nabla^\alpha\nabla^\lambda\varphi\nabla^\beta\nabla_\lambda\varphi -3\mathcal{W}_{\alpha\rho\beta\sigma}R^{\rho\sigma}\nabla^\alpha\varphi\nabla^\beta\varphi\nabla^\lambda\varphi\nabla_\lambda\varphi \right. \nonumber \\ 
&& \ \ \ \ \ \ \ \ \left. { + \frac{3}{2}R_{\mu\alpha\beta\gamma}R_{\nu}^{\ \ \alpha\beta\gamma}\nabla^{\mu}\varphi\nabla^{\nu}\varphi\nabla^{\lambda}\varphi\nabla_{\lambda}\varphi} \right] \nonumber \\
&& + 2\frac{M_{\rm Pl}}{M^3}c_GG_{\alpha\beta}\nabla^\alpha\nabla^\beta\varphi.
\eq
\end{widetext}

The usual equations presented in the literature (\cite{PhysRevD.79.084003, Appleby:2011aa}, e.g.) are related to ours via the following Riemann tensor expansion

\bq\label{Riemann-expansion}
R_{\mu\nu\alpha\beta} = && \frac{1}{2}\left(g_{\mu\alpha}R_{\nu\beta} + g_{\nu\beta}R_{\mu\alpha} - g_{\mu\beta}R_{\nu\alpha} - g_{\nu\alpha}R_{\mu\beta}\right) \nonumber \\
&& + \mathcal{W}_{\mu\nu\alpha\beta} - \frac{1}{6}R\left( g_{\mu\alpha}g_{\nu\beta} - g_{\mu\beta}g_{\nu\alpha}\right),
\eq
which cancels some of the terms originally derived in \cite{PhysRevD.79.084003}. {In Eq. (\ref{Full-EOM}) we did not write the term proportional to $c_5R_{\mu\alpha\beta\gamma}R_{\nu}^{\ \ \alpha\beta\gamma}$ using Eq. (\ref{Riemann-expansion}) as in this particular case the expansion would make the equations longer.}

\section{Alternative Derivation of the Perturbed Equations}

\label{alternative_derivation}

In this appendix we present an alternative derivation of the perturbed equations in Galileon gravity. This method requires only the knowledge of the Galileon equation of motion and the assumption that all the field equations do not contain derivatives higher than second order, the latter being satisfied by the theory of Galileon gravity by definition. 

If the above requisites are satisfied, then it is easier to derive the perturbed components of the Galileon energy-momentum tensor using the new method rather than from the complicated Galileon Lagrangian. In the latter case, one has to first derive the full expressions of the energy-momentum tensor (see Appendix~\ref{Appendix A}), which itself could be a considerable amount of work.

The spirit of this derivation follows the general method introduced in \citep{Skordis:2008vt} and generalised later by \citep{Ferreira:2010sz,Baker:2011jy,Zuntz:2011aq,Baker:2012zs}. However, here we work within the framework of covariant and gauge-invariant perturbations, and consequently the mathematical description looks different from those works. 

To lighten the notation, in this appendix we neglect the superscript $^G$ in the dynamical quantities for the Galileon field.

\subsection{The Method}

As we have seen above, the quantities $\rho, p, q_{\mu}$ and $\pi_{\mu\nu}$ have contributions from both normal matter and the Galileon field $\varphi$. Here, let us first look at the most general forms that $\rho, p, q_{\mu}$ and $\pi_{\mu\nu}$ for the Galileon field can take. The arguments are as follows:
\begin{enumerate}
\item Eq.~(\ref{spatial curvature}) contains time derivatives up to first order (in $\theta$) and spatial derivatives up to second order (in $\hat{R}$). If we want to keep this property, the Galileon energy density $\rho$ can contain $\theta$, $\hat{R}$, $\dot{\varphi}$ and $\hat{\Box}\varphi$, but not their time derivatives or gradients. It cannot contain quantities such as $\grad3^\mu\grad3^\nu\sigma_{\mu\nu}$, which involve higher order derivatives.  If it contains $\grad3\cdot A$, then according to Eq.~(\ref{conservation1}) $p$ must contain $(\grad3\cdot A)^{\cdot}$ which involves third-order derivative and hence it is not allowed.
\item According to Eq.~(\ref{conservation1}), the Galileon pressure $p$ can contain $\dot{\theta}, \theta, \dot{\varphi}, \hat{R}, \ddot{\varphi}$. It can also contain $\grad3\cdot A$ without changing the structure of Eq.~(\ref{p1}). Quantities such as $\dot{\hat{R}}, (\grad3\cdot A)^{\cdot}$ and $\grad3^\mu\grad3^\nu\sigma_{\mu\nu}$ are not allowed as they contain higher-order derivatives. 
\item The Galileon field peculiar velocity $q_{\mu}$  can contain $\grad3_\mu\theta$, $\grad3^\nu\sigma_{\mu\nu}, \grad3^{\nu}\varpi_{\mu\nu}, \grad3_\mu\varphi$ and $\grad3_\mu\dot{\varphi}$, but not their time and spatial derivatives. If it contains $A_{\mu}$, then Eq.~(\ref{conservation2}) cannot hold without involving derivatives higher than second order. It cannot contain $\grad3_\mu\hat{R}$ because this has third order derivatives. $\ddot{\varphi}$ and $\dot{\theta}$ are not allowed because otherwise either $p$ or $\pi_{\mu\nu}$ has to contain third order time derivatives, according to Eq.~(\ref{conservation2}).
\item The Galileon anisotropic stress tensor $\pi_{\mu\nu}$ can contain $\sigma_{\mu\nu}$, $\dot{\sigma}_{\mu\nu}, \mathcal{E}_{\mu\nu}, \grad3_{\langle\mu}\grad3_{\nu\rangle}$ and $\grad3_{\langle\mu}A_{\nu\rangle}$, but not their time and spatial derivatives. 
\end{enumerate}

Based on the above analysis, we can decide which terms can appear in the expressions of $\rho, p, q_{\mu}, \pi_{\mu\nu}$ for the Galileon field. More explicitly, up to first order in linear perturbations, we have
\begin{eqnarray}\label{eq:rho}
\label{eq:ab-rho}\rho &=& A_\rho\dot{\varphi}^{a}\theta^{b}+B_\rho\dot{\varphi}^{a}\theta^{b-2}\hat{R}+C_\rho\dot{\varphi}^{a-1}\theta^{b-1}\hat{\Box}\varphi,\\
\label{eq:ab-p}p &=& A_p\ddot{\varphi}\dot{\varphi}^{a-1}\theta^{b-1}+B_p\dot{\varphi}^{a}\dot{\theta}\theta^{b-2}+C_p\dot{\varphi}^{a}\theta^{b}\nonumber\\
&&+D_p\ddot{\varphi}\dot{\varphi}^{a-2}\theta^{b-2}\hat{\Box}\varphi + E_p\dot{\varphi}^{a-1}\dot{\theta}\theta^{b-3}\hat{\Box}\varphi\nonumber\\
&&+F_p\dot{\varphi}^{a-1}\theta^{b-1}\hat{\Box}\varphi + G_p\ddot{\varphi}\dot{\varphi}^{a-1}\theta^{b-3}\grad3\cdot A\nonumber\\
&&+H_p\dot{\varphi}^{a}\dot{\theta}\theta^{b-4}\grad3\cdot A+I_p\dot{\varphi}^{a}\theta^{b-2}\grad3\cdot A\nonumber\\
&&+J_p\ddot{\varphi}\dot{\varphi}^{a-1}\theta^{b-3}\hat{R}+K_p\dot{\varphi}^{a}\dot{\theta}\theta^{b-4}\hat{R}\nonumber\\
&&+L_p\dot{\varphi}^{a}\theta^{b-2}\hat{R},\\
\label{eq:ab-q}q_\mu &=& A_q\dot{\varphi}^{a-1}\theta^{b-1}\grad3_\mu\dot{\varphi} + B_q\dot{\varphi}^{a-1}\theta^{b}\grad3_\mu\varphi\nonumber\\
&&+C_q\dot{\varphi}^a\theta^{b-2}\grad3_\mu\theta + D_q\dot{\varphi}^a\theta^{b-2}\grad3^\nu\sigma_{\mu\nu}\nonumber\\
&&+E_q\dot{\varphi}^a\theta^{b-2}\grad3^\nu\varpi_{\mu\nu},\\
\label{eq:ab-pi}\pi_{\mu\nu} &=& A_\pi\dot{\varphi}^a\theta^{b-2}\mathcal{E}_{\mu\nu} + B_\pi\ddot{\varphi}\dot{\varphi}^{a-1}\theta^{b-3}\mathcal{E}_{\mu\nu}\nonumber\\
&&+C_\pi\dot{\varphi}^a\dot{\theta}\theta^{b-4}\mathcal{E}_{\mu\nu} + D_\pi\dot{\varphi}^a\theta^{b-2}\dot{\sigma}_{\mu\nu}\nonumber\\
&&+E_\pi\ddot{\varphi}\dot{\varphi}^{a-1}\theta^{b-3}\dot{\sigma}_{\mu\nu}+F_\pi\dot{\varphi}^a\dot{\theta}\theta^{b-4}\dot{\sigma}_{\mu\nu}\nonumber\\
&&+G_\pi\dot{\varphi}^a\theta^{b-1}\sigma_{\mu\nu} + H_\pi\ddot{\varphi}\dot{\varphi}^{a-1}\theta^{b-2}\sigma_{\mu\nu}\nonumber\\
&&+I_\pi\dot{\varphi}^a\dot{\theta}\theta^{b-3}\sigma_{\mu\nu}  + J_\pi\dot{\varphi}^a\theta^{b-2}\grad3_{\langle\mu}A_{\nu\rangle}\nonumber\\
&&+K_\pi\ddot{\varphi}\dot{\varphi}^{a-1}\theta^{b-3}\grad3_{\langle\mu}A_{\nu\rangle}+L_\pi\dot{\varphi}^a\dot{\theta}\theta^{b-4}\grad3_{\langle\mu}A_{\nu\rangle}\nonumber\\
&&+M_\pi\ddot{\varphi}\dot{\varphi}^{a-2}\theta^{b-2}\grad3_{\langle\mu}\grad3_{\nu\rangle}\varphi + N_\pi\dot{\varphi}^{a-1}\theta^{b-1}\grad3_{\langle\mu}\grad3_{\nu\rangle}\varphi\nonumber\\
&&+O_\pi\dot{\varphi}^{a-1}\dot{\theta}\theta^{b-3}\grad3_{\langle\mu}\grad3_{\nu\rangle}\varphi,
\end{eqnarray}
in which $A_{\rho, p, q, \pi}, B_{\rho, p, q, \pi}, \cdots$ are constant coefficients and $a, b$ are dimensionless constant power indices. Note that to write down the above equations we have used the fact that
\begin{enumerate}
\item all terms in the expressions must have the same mass dimension and
\item the power of $\varphi$ (with $\dot{\varphi}, \ddot{\varphi}$ and $\hat{\Box}\varphi$ counted in) must be the same in all terms,
\end{enumerate}
which must be true if the dynamical quantities are to be derived from the Lagrangian densities $\mathcal{L}_1$--$\mathcal{L}_5$ that are specified in Eq.~(\ref{L's}). 

When using the above expressions, we require that all terms must not contain negative powers of $\theta$ (which will never appear when varying $\mathcal{L}_{1-5}$ with respect to the metric $g_{\mu\nu}$). For example, if $b=3$, then $K_p$ should be set to zero.

\subsection{Application of the Method: the $c_4$ Term}

Here we illustrate the application of our method for the particular case of the $c_4$ term. The Lagrangian $\mathcal{L}_4$ is sufficiently complicated to highlight how much simpler this method can be. For this term, we know from the background expression of the energy density (or equivalently the Galileon equation of motion) that $a=4, b=2$, and so we can write
\begin{eqnarray}
\rho &=& A_\rho\dot{\varphi}^{4}\theta^{2}+B_\rho\dot{\varphi}^{4}\hat{R}+C_\rho\dot{\varphi}^{3}\theta\hat{\Box}\varphi,\\
p &=& A_p\ddot{\varphi}\dot{\varphi}^{3}\theta+B_p\dot{\varphi}^{4}\dot{\theta}+C_p\dot{\varphi}^{4}\theta^{2}+D_p\ddot{\varphi}\dot{\varphi}^{2}\hat{\Box}\varphi\nonumber\\
&&+F_p\dot{\varphi}^{3}\theta\hat{\Box}\varphi+I_p\dot{\varphi}^{4}\grad3\cdot A+L_p\dot{\varphi}^{4}\hat{R},\\
q_\mu &=& A_q\dot{\varphi}^{3}\theta\grad3_\mu\dot{\varphi} + B_q\dot{\varphi}^{3}\theta^{2}\grad3_\mu\varphi\nonumber\\
&&+C_q\dot{\varphi}^4\grad3_\mu\theta + D_q\dot{\varphi}^4\grad3^\nu\sigma_{\mu\nu}+E_q\dot{\varphi}^4\grad3^\nu\varpi_{\mu\nu},\\
\pi_{\mu\nu} &=& A_\pi\dot{\varphi}^4\mathcal{E}_{\mu\nu} +D_\pi\dot{\varphi}^4\dot{\sigma}_{\mu\nu}
+G_\pi\dot{\varphi}^4\theta\sigma_{\mu\nu}\nonumber\\ 
&&+ H_\pi\ddot{\varphi}\dot{\varphi}^3\sigma_{\mu\nu}+J_\pi\dot{\varphi}^4\grad3_{\langle\mu}A_{\nu\rangle}\nonumber\\
&&+M_\pi\ddot{\varphi}\dot{\varphi}^2\grad3_{\langle\mu}\grad3_{\nu\rangle}\varphi + N_\pi\dot{\varphi}^{3}\theta\grad3_{\langle\mu}\grad3_{\nu\rangle}\varphi.
\end{eqnarray}
Substituting these into the conservation equations (\ref{conservation1}, \ref{conservation2}), we find
\begin{widetext}
\begin{eqnarray}
\label{eq:c4_cons}(4A_\rho+A_p)\ddot{\varphi}\dot{\varphi}^3\theta^2 + (2A_\rho+B_p)\dot{\varphi}^4\dot{\theta}\theta + (A_\rho+C_p)\dot{\varphi}^4\theta^3 + \left[4B_\rho\ddot{\varphi}\dot{\varphi}^3+\left(\frac{1}{3}B_\rho+L_p\right)\dot{\varphi}^4\theta\right]\hat{R}\nonumber\\
+\left[(3C_\rho+D_p)\ddot{\varphi}\dot{\varphi}^2\theta+C_\rho\dot{\varphi}^3\dot{\theta}+\left(\frac{2}{3}A_q+B_q+F_p+C_\rho\right)\dot{\varphi}^3\theta^2\right]\hat{\Box}\varphi + (A_q+C_\rho)\dot{\varphi}^3\theta\left(\hat{\Box}\varphi\right)^{\cdot}\nonumber\\
+\left[\frac{4}{3}B_\rho+C_q\right]\dot{\varphi}^2\hat{\Box}\theta+\left[A_q+I_p-\frac{4}{3}B_\rho\right]\dot{\varphi}^4\theta\grad3\cdot A+(D_q-2B_\rho)\dot{\varphi}^4\grad3^\mu\grad3^\nu\sigma_{\mu\nu}+(E_q-2B_\rho)\dot{\varphi}^4\grad3^\mu\grad3^\nu\varpi_{\mu\nu} &=& 0,\\
\label{eq:c4_cont} (A_q-A_p)\dot{\varphi}^3\theta\left(\grad3_\mu\dot{\varphi}\right)^{\cdot} + \left[3(A_q-A_p)\ddot{\varphi}\dot{\varphi}^2\theta+(A_q-4B_p)\dot{\varphi}^3\dot{\theta}+\left(B_q+\frac{4}{3}A_q-4C_p-\frac{1}{3}A_p\right)\dot{\varphi}^3\theta^2\right]\grad3_\mu\dot{\varphi}\nonumber\\
+B_q\left[3\ddot{\varphi}\dot{\varphi}^2\theta^2+\dot{\varphi}^3\theta^3+2\dot{\varphi}^3\dot{\theta}\theta\right]\grad3_\mu\varphi + (A_\pi-12L_p)\dot{\varphi}^4\grad3^\nu\mathcal{E}_{\mu\nu} + (D_q+D_\pi-6L_p)\dot{\varphi}^4\left(\grad3^\nu\sigma_{\mu\nu}\right)^{\cdot}\nonumber\\ 
+ (E_q-6L_p-J_\pi)\dot{\varphi}^4\left(\grad3^\nu\varpi_{\mu\nu}\right)^{\cdot} + \left[(4D_q+H_\pi)\ddot{\varphi}\dot{\varphi}^3+\left(\frac{4}{3}D_q+\frac{1}{3}D_\pi+G_\pi-4L_p\right)\dot{\varphi}^4\theta\right]\grad3^\nu\sigma_{\mu\nu}\nonumber\\
+\left[(4E_q+M_\pi)\ddot{\varphi}\dot{\varphi}^3+\left(\frac{4}{3}E_q+N_\pi-4L_p-J_\pi\right)\dot{\varphi}^4\theta\right]\grad3^\nu\varpi_{\mu\nu} 
+(A_\rho-B_q+C_p)\dot{\varphi}^4\theta^2A_\mu\nonumber\\
+(C_q-B_p)\dot{\varphi}^4\left(\grad3_\mu\theta\right)^{\cdot} + \left[(4C_q-A_p)\ddot{\varphi}\dot{\varphi}^3+\left(\frac{4}{3}C_q-2C_p-\frac{1}{3}B_p\right)\dot{\varphi}^4\theta\right]\grad3_\mu\theta\nonumber\\
-\left[I_p-4L_p-\frac{2}{3}J_\pi\right]\dot{\varphi}^4\grad3_\mu\grad3\cdot A
- \left[\left(D_p-\frac{2}{3}M_\pi\right)\ddot{\varphi}\dot{\varphi}^2+\left(F_p-\frac{2}{3}N_\pi\right)\dot{\varphi}^3\theta\right]\grad3_\mu\hat{\Box}\varphi &=& 0.\ \ \ \
\end{eqnarray}
\end{widetext}
From the background expression of $\rho$ and $p$ (or equivalently the background Galileon equation of motion together with the energy conservation equation) for the $c_4$ term, we find
\begin{eqnarray}
A_\rho = \frac{5}{2}\lambda,\ A_p = -4\lambda,\ B_p = -\lambda,\ C_p = -\frac{1}{2}\lambda,
\end{eqnarray}
in which $\lambda\equiv c_4/M^6$. This can be done by equating the first three terms of Eq.~(\ref{eq:c4_cons}) to the background Galileon equation of motion
\begin{eqnarray}
3\ddot{\varphi}\dot{\varphi}\theta^2+2\dot{\varphi}^2\dot{\theta}\theta+\dot{\varphi}^2\theta^3 &=& 0,
\end{eqnarray}
which can also be used to eliminate the terms containing $\grad3_\mu\varphi$ in Eq.~(\ref{eq:c4_cont}). 

Because we have already used the Galileon equation of motion in Eq.~(\ref{eq:c4_cont}), the remaining terms on the left-hand side of this equation must cancel amongst themselves. In addition, for Eq.~(\ref{eq:c4_cons}) to not contain higher-order derivatives, we must set the coefficients of $\left(\hat{\Box}\varphi\right)^\cdot$, $\hat{\Box}\theta$, $\grad3^\mu\grad3^\nu\sigma_{\mu\nu}$ and $\grad3^\mu\grad3^\nu\varpi_{\mu\nu}$ to zero. This gives us
\begin{eqnarray}
C_\rho = -D_p = -A_q = 4\lambda,\  C_q = -I_p = -\lambda,\nonumber\\
B_q = 2\lambda,\ H_\pi = M_\pi = -6\lambda,\ D_q=E_q=2B_\rho=\frac{3}{2}\lambda,\nonumber
\end{eqnarray}
and
\begin{eqnarray}
L_p &=& \frac{1}{6}D_\pi + \frac{1}{4}\lambda,\nonumber\\
A_\pi &=& 2D_\pi+3\lambda,\nonumber\\
G_\pi &=& \frac{1}{3}D_\pi-\lambda,\nonumber\\
N_\pi &=& -\frac{1}{3}D_\pi-\lambda,\nonumber\\
F_\pi &=& -\frac{2}{9}D_\pi-\frac{2}{3}\lambda.
\end{eqnarray}

Unfortunately, some coefficients cannot be fixed unambiguously, and here we have expressed all those coefficients in terms of $D_\pi$. This indicates that perhaps the Galileon model is not the only one which gives the perturbed energy-momentum tensor as in Eqs.~(\ref{eq:ab-rho},\ref{eq:ab-p},\ref{eq:ab-q},\ref{eq:ab-pi}). To solve this problem, we can use the perturbed Galileon equation of motion to fix the free parameter. Of course, this does not necessarily mean that we have to write down the full perturbed equation of motion. Indeed, we only need to know the coefficient $L_p$ or $F_p$.

The Galileon equation of motion can be read from the remaining terms of Eq.~(\ref{eq:c4_cons}), from which we find that the ratio of the coefficients of $\ddot{\varphi}\dot{\varphi}^3\theta^2$ and $\dot{\varphi}^4\theta\hat{R}$ is $1/12+D_\pi/(36\lambda)$. On the other hand, the value of this ratio can also be easily calculated by explicitly perturbing the Galileon equation of motion where we find it to be $1/18$. As a result, $D_\pi=-\lambda$ and all the coefficients are now fixed. The components of the energy-momentum tensor of the $c_4$ term are:

\begin{widetext}
\begin{eqnarray}
\label{eq:c4_rho}\rho &=& \frac{c_4}{M^6}\left[\frac{5}{2}\dot{\varphi}^{4}\theta^{2}+\frac{3}{4}\dot{\varphi}^{4}\hat{R}+4\dot{\varphi}^3\theta\hat{\Box}\varphi\right],\\
\label{eq:c4_p}p &=& \frac{c_4}{M^6}\left[-4\ddot{\varphi}\dot{\varphi}^3\theta-\dot{\varphi}^{4}\dot{\theta}-\frac{1}{2}\dot{\varphi}^{4}\theta^{2}-4\ddot{\varphi}\dot{\varphi}^2\hat{\Box}\varphi-\frac{4}{9}\dot{\varphi}^3\theta\hat{\Box}\varphi +\dot{\varphi}^{4}\grad3\cdot A+\frac{1}{12}\dot{\varphi}^{4}\hat{R}\right],\\
\label{eq:c4_q}q_\mu &=& \frac{c_4}{M^6}\left[-4\dot{\varphi}^3\theta\grad3_\mu\dot{\varphi} + 2\dot{\varphi}^3\theta^{2}\grad3_\mu\varphi-\dot{\varphi}^4\grad3_\mu\theta + \frac{3}{2}\dot{\varphi}^4\grad3^\nu(\sigma_{\mu\nu}+\varpi_{\mu\nu})\right],\\
\label{eq:c4_pi}\pi_{\mu\nu} &=& \frac{c_4}{M^6}\left[\dot{\varphi}^4\mathcal{E}_{\mu\nu}-\dot{\varphi}^4\dot{\sigma}_{\mu\nu}+\dot{\varphi}^4\grad3_{\langle\mu}A_{\nu\rangle}-\frac{4}{3}\dot{\varphi}^4\theta\sigma_{\mu\nu}-6\ddot{\varphi}\dot{\varphi}^3\sigma_{\mu\nu}-6\ddot{\varphi}\dot{\varphi}^2\grad3_{\langle\mu}\grad3_{\nu\rangle}\varphi-\frac{2}{3}\dot{\varphi}^3\theta\grad3_{\langle\mu}\grad3_{\nu\rangle}\varphi\right],
\end{eqnarray}
and the fully perturbed Galileon equation of motion becomes
\begin{eqnarray}\label{eq:c4_eom}
0 &=& 6\ddot{\varphi}\dot{\varphi}^2\theta^2 + 4\dot{\varphi}^3\dot{\theta}\theta + 2\dot{\varphi}^3\theta^3 - 4\dot{\varphi}^3\theta\grad3\cdot A+\left[3\ddot{\varphi}+\frac{1}{3}\dot{\varphi}\theta\right]\dot{\varphi}^2\hat{R} + \left[8\ddot{\varphi}\dot{\varphi}\theta+4\dot{\varphi}^2\dot{\theta}+\frac{26}{9}\dot{\varphi}^2\theta^2\right]\hat{\Box}\varphi,
\end{eqnarray}
\end{widetext}
which is in agreement with the $c_4$ terms in Eqs.~(\ref{perturbed1} - \ref{perturbed4}) and Eq.~(\ref{perturbed EoM}), respectively.

We have applied the same method to all other terms, and for all of them the resulted equations agree with Eqs.~(\ref{perturbed1} - \ref{perturbed4}) and Eq.~(\ref{perturbed EoM})\footnote{For the cases of $c_2$ and $c_3$ terms, all coefficients can be fixed unambiguously}. Note that in this new method the different terms of the Galileon field can be worked out in a unified way, which further reduces the computational effort. With certain modifications, the method should be applicable to the generalised Galileon model \citep{DeFelice:2010nf} as well.


\bibliography{text.bib}

\end{document}